\documentclass[journal]{IEEEtran}
\IEEEoverridecommandlockouts
\usepackage{cite}
\usepackage{amsmath,amssymb,amsfonts}
\usepackage{algorithmic}
\usepackage{graphicx}
\usepackage{textcomp}
\usepackage{xcolor}
\usepackage{relsize}
\usepackage{booktabs}
\usepackage{multicol}
\usepackage{multirow}
\usepackage{amssymb}

\makeatletter
\let\MYcaption\@makecaption
\makeatother
\usepackage{caption}
\usepackage[font=footnotesize]{subcaption}
\makeatletter
\let\@makecaption\MYcaption
\makeatother

\DeclareSymbolFont{EULE}{U}{eur}{m}{n}
\DeclareMathAlphabet{\matheule}{U}{eur}{m}{n}

\makeatletter
\def\argmin{\mathop{\operator@font argmin}}
\def\cs{\mkern 1mu}
\def\clap#1{\hbox to 0pt{\hss#1\hss}}

\def\mathrlap{\mathpalette\mathrlapinternal}

\def\mathrlapinternal#1#2{%
    \rlap{$\mathsurround=0pt#1{#2}$}}

\makeatother

\newcommand*{\bigs}[1]{\vcenter{\hbox{\scalebox{2.1}{\ensuremath#1}}}}

\def\ve#1{{\mathchoice{\mbox{\boldmath$\displaystyle #1$}}%
              {\mbox{\boldmath$\textstyle #1$}}%
              {\mbox{\boldmath$\scriptstyle #1$}}%
              {\mbox{\boldmath$\scriptscriptstyle #1$}}}}

\def\vee#1{\mathrlap{\ve{#1}}\phantom{\ve{#1}}}              

\def\defeq{\stackrel{\mbox{\tiny def}}{=}}

\DeclareMathSymbol{\pdf}{\mathalpha}{EULE}{"66}
\DeclareMathSymbol{\cdf}{\mathalpha}{EULE}{"46}
\DeclareMathSymbol{\GammaF}{\mathalpha}{operators}{"00}
\DeclareMathSymbol{\ThetaF}{\mathalpha}{operators}{"02}
\DeclareMathSymbol{\cf}{\mathalpha}{operators}{"08}
\DeclareMathSymbol{\step}{\mathalpha}{EULE}{"22}
\DeclareMathSymbol{\dirac}{\mathalpha}{EULE}{"0E}
\def\sinc{\mathop{\mathrm{sinc}}}

\newtheorem{theorem}{Theorem}[section]

\newtheorem{lemma}[theorem]{Lemma}

\usepackage{microtype}
\makeatletter
\def\ps@IEEEtitlepagestyle{%
  \def\@oddfoot{\mycopyrightnotice}%
  \def\@oddhead{\hbox{}\@IEEEheaderstyle\leftmark\hfil\thepage}\relax
  \def\@evenhead{\@IEEEheaderstyle\thepage\hfil\leftmark\hbox{}}\relax
  \def\@evenfoot{}%
}
\def\mycopyrightnotice{%
  \begin{minipage}{\textwidth}
  \centering \scriptsize
  © 2022 IEEE. Personal use of this material is permitted. Permission
from IEEE must be obtained for all other uses, in any current or future
media, including reprinting/republishing this material for advertising or
promotional purposes, creating new collective works, for resale or
redistribution to servers or lists, or reuse of any copyrighted
component of this work in other works.
  \end{minipage}
}
\makeatother

\begin{document}

\title{Indoor Localization with Robust Global Channel Charting: A Time-Distance-Based Approach}

	

\author{
Maximilian Stahlke, 
    George Yammine, 
	Tobias Feigl, 
	Bjoern M.\ Eskofier, 
	Christopher Mutschler 
\thanks{This work has been submitted to the IEEE for possible publication. Copyright may be transferred without notice, after which this version may no longer be accessible.}
\thanks{This work was supported by the Fraunhofer Lighthouse project ``6G SENTINEL'' and by the Federal Ministry of Education and Research of Germany in the programme of ``Souver\"an. Digital. Vernetzt.'' joint project 6G-RIC (16KISK020K).}
\thanks{M.\ Stahlke, G.\ Yammine, T.\ Feigl and C.\ Mutschler are with Fraunhofer IIS, Fraunhofer Institute for Integrated Circuits IIS, Division Positioning and Networks, 90411 Nuremberg, Germany (email: maximilian.stahlke@iis.fraunhofer.de; george.yammine@iis.fraunhofer.de; tobias.feigl@iis.fraunhofer.de; christopher.mutschler@iis.fraunhofer.de).}
\thanks{B.\ M.\ Eskofier is with the Department Artificial Intelligence
in Biomedical Engineering (AIBE), Friedrich-Alexander-Universit\"at
Erlangen-N\"urnberg (FAU), 91052 Erlangen, Germany (email: bjoern.eskofier@fau.de).}
}



\maketitle

\begin{abstract}
Fingerprinting-based positioning significantly improves the indoor localization performance in non-line-of-sight-dominated areas. However, its deployment and maintenance is cost-intensive as it needs ground-truth reference systems for both the initial training and the adaption to environmental changes. In contrast, channel charting (CC) works without explicit reference information and only requires the spatial correlations of channel state information (CSI). While CC has shown promising results in modelling the geometry of the radio environment, a deeper insight into CC for localization using multi-anchor large-bandwidth measurements is still pending. 

We contribute a novel distance metric for time-synchronized single-input/single-output CSIs that approaches a linear correlation to the Euclidean distance. This allows to learn the environment's global geometry without annotations. To efficiently optimize the global channel chart we approximate the metric with a Siamese neural network. This enables full CC-assisted fingerprinting and positioning only using a linear transformation from the chart to the real-world coordinates. We compare our approach to the state-of-the-art of CC on two different real-world data sets recorded with a 5G and UWB radio setup. Our approach outperforms others with localization accuracies of 0.69\,m for the UWB and 1.4\,m for the 5G setup. We show that CC-assisted fingerprinting enables highly accurate localization and reduces (or eliminates) the need for annotated training data.

\end{abstract}


\section{Introduction}



\IEEEPARstart{I}{ndoor} localization techniques are a key enabler for several downstream tasks in health care, industrial production or networking \cite{laoudias2018survey}. Although several localization techniques based on camera images \cite{wu2018image}, lidar sensors \cite{elhousni2020survey} or visible light communication \cite{rahman2020recent} already exist, radio-based localization is still one of the most promising technologies \cite{saeed2019state}. Approaches such as angle-of-arrival (AoA) \cite{pang2020aoa} or time-of-arrival (ToA) \cite{gifford2020impact} localization can achieve accuracies in the centimeter range given a line-of-sight (LoS) path between transmitters and receivers. However, this requirement can often not be met due to the complexity of indoor environments. While there are methods that can mitigate the non-line-of-sight (NLoS) effects \cite{christopher2021characterizing} by identification and exclusion \cite{stahlke2020nlos} or error estimation \cite{stahlke2021estimating}, a majority of radio units still need LoS for localization. To anyway achieve highly accurate localization in NLoS-dominated areas, fingerprinting methods can be used. Fingerprinting exploits the channel state information (CSI) collected in a radio environment. The CSI contains location-specific radio information caused by reflections, scattering and absorptions~\cite{niitsoo2019deep, stahlke20225g, liu2017toward, de2020csi, widmaier2019towards}. To train a fingerprinting model, CSI measurements have to be labeled using a ground-truth reference system, which is often very expensive, while another common problem of fingerprinting is that environmental changes can alter the location-specific fingerprints, making regular updates inevitable, which again requires labeled data~\cite{stahlke20225g, widmaier2019towards}. 

Recently, a new concept called channel charting (CC) has been proposed. CC generates a chart on collected CSI measurements to reflect the (local) geometry of the environment in an unsupervised manner, i.e., no ground-truth labels are needed \cite{studer2018channel}. CC exploits the fact that CSI measurements are spatially correlated. That allows to obtain a chart in the spatial or angular domain. The obtained chart does not represent the global coordinates of the environment. Instead, it resembles only the relative geometry of the radio environment. To use CC for positioning, some semi-supervised approaches \cite{lei2019siamese, ferrand2021triplet, deng2021network, zhang2021semi} use few labeled data points to transform the channel chart into the physical domain and enhance its consistency. 
It is mainly used to leverage several downstream tasks for multiple-input/multiple-output (MIMO) communications, e.g., pilot assignment \cite{ribeiro2020channel}, UE grouping \cite{al2021adaptive}, radio resource management \cite{al2020multipoint} or beam forming \cite{ponnada2021location, ponnada2021best, kazemi2021channel}.

However, all of the existing work on CC is based solely on MIMO systems with either single or multiple unsynchronized base stations with mostly less than 50\,MHz bandwidth. Due to the low bandwidth and the lack of synchronization, the networks are less suited for localization and are rather optimized for communication. This renders current CC approaches inappropriate for high precision positioning tasks. Thus, we investigate the abilities of CC for high-precision indoor localization employing time-synchronized single-input/single-output (SISO) radio systems with high bandwidth. We contribute a novel geodesic CSI distance metric that provides a linear correlation to the Euclidean distance. A Siamese network generates channel charts which reflect the global instead of only the local geometry to enable CC-assisted fingerprinting by only applying an affine transformation after optimization. We compare all the relevant methods for CC with ours on two real-world data sets using UWB and 5G radio systems outperforming the state-of-the-art with a localization accuracy of 0.69\,m for the UWB and 1.4\,m for the 5G setup.

The remainder of this article is structured as follows. Section~\ref{section:related_work}
discusses related work. Next, Section~\ref{section:method} provides details about CIR distances, geodesic distances, how we approximate them using a Siamese neural network, and how we process the data. Section~\ref{section:experimental_setup} describes our experimental setup. The numerical results are presented in  Section~\ref{section:evaluation} and discussed in Section~\ref{section:discussion}. Section~\ref{section:conclusion} concludes. 


\section{Related Work}
\label{section:related_work}

\noindent Studer et al.~\cite{studer2018channel} first described the concept of CC for multi-antenna systems, e.g., MIMO architectures, to embed the local radio geometry. As channel charts only yield poor geometries using a single base station (BS), Deng et al.~\cite{deng2018multipoint} proposed a multi-point CC approach that fuses the channel information of multiple BSs to achieve a better spatial consistency of the channel charts. CC typically consists of two components, (i) a distance metric, which reflects the Euclidean distance between two channel measurements and (ii) the dimensionality reduction, which encodes the high dimensional channel information into a 2D embedding reflecting the position.

As a distance metric, Studer et al.~\cite{studer2018channel} propose to use the free-space path loss between two measurements, which is proportional to the Euclidean distance if the translation is collinear and the path-loss coefficient is constant. Moreover, Magoarou et al.~\cite{le2021efficient} enhanced the proposed metric by making it insensitive to fast fading effects by putting the channel measurements into phase. As the distance metric is only valid for collinear measurements, Agostini et al.~\cite{agostini2020channel} proposed a grouping of collinear measurements by exploiting the angular information encoded in the CSI measurements of fully digital MIMO setups. To achieve channel charts with higher quality, in fully digital MIMO setups, multipath components (MPCs) can be extracted using MUltiple Signal Identification Classification (MUSIC) \cite{schmidt1986multiple}, which allows to cluster MPCs and apply the path-loss model for every component providing an improvement of the original metric. However, MUSIC needs a known number of MPCs, which is often difficult to obtain especially under low signal-to-noise ratio (SNR) situations \cite{spielman1986music}. Moreover, a data association of MPCs after extraction is necessary, which is non-trivial especially in complex environments with dense multipath propagation with varying MPCs.

For indoor localization, (synchronized) SISO radio setups are usually preferred as multiple base stations are needed. MIMO-capable base stations, in contrast, are more expensive than their SISO counterparts, due to their higher hardware complexity mostly useful for communication tasks.
Compared to MIMO setups, SISO base stations have a simple hardware concept as they have only one receiving antenna but therefore cannot estimate any AoA information for positioning. Therefore, ToF or TDoA estimations are typically used for positioning \cite{gifford2020impact}. 
However, this also means that for SISO setups, the angular-based distance metrics presented in the literature cannot be used. We therefore contribute a novel distance metric based on raw CSI measurements in the time-domain exploiting ToF and TDoA information dedicated for low cost SISO positioning systems. 

The second component of CC is its dimensionality reduction, which can be accomplished using both parametric approaches and non-parametric approaches. Studer et al.~\cite{studer2018channel} investigated non-parametric approaches such as principal component analysis (PCA) or Sammon mapping to CC. Ponnada et al.~\cite{ponnada2019out} proposed an approach based on Laplacian-Eigenmaps with an extension to prediction also on unseen data, which is the main intention of CC. This is also done implicitly by mapping the high-dimensional channel information to 2D embeddings using parametric deep learning (DL) approaches. Methods such as autoencoders \cite{geng2020multipoint}, constrained autoencoders \cite{huang2019improving}, Siamese networks \cite{lei2019siamese}, and triplets \cite{ferrand2020triplet, ferrand2021triplet, rappaport2021improving, euchner2022improving} have shown very promising results in modelling the local geometry of the channel information. Also, a combined algorithm is proposed, which first compresses the CSI using a convolutional autoencoder to efficiently apply uniform manifold approximation and projection (UMAP) on the lower-dimensional representation of the CSI \cite{agostini2022not}. While CC only models the local geometry of the area, there are also semi-supervised approaches that learn the mapping from the local to the global coordinate frame. This can either be achieved by an affine transformation after optimizing the local map \cite{pihlajasalo2020absolute}, or as a constraint in the optimization loss \cite{lei2019siamese, ferrand2021triplet, deng2021network, zhang2021semi}. While the objective of the aforementioned approach is to minimize the local distance between adjacent points, Magoarou et al.~\cite{le2021efficient} used Isomap to optimize the channel chart between all CSIs on global distances, which are created as sum of local distances on the shortest path between two points. This allows to generate a chart with global instead of only local similarity. 

This work builds upon the idea of Isomap and proposes to leverage geodesic distances in a Siamese network to learn a globally consistent channel chart that enables efficient and accurate location prediction on unseen data. We theoretically and practically show that our metric has a high linear correlation to the Euclidean distance enabling channel charting for SISO radio systems for CC-assisted fingerprinting. We compare the performance of all relevant CC approaches w.r.t.\ global and local similarity on two real-world data sets based on UWB and 5G radio setups.

\section{Method}
\label{section:method}

\noindent To estimate a globally consistent channel chart, we need both a locally and globally valid distance metric for CSI. First, we derive our local distance in Section~\ref{chap:cir_dist} and our approximation of the global distance in Section~\ref{chap:geo_dist}. Next, we describe the architecture of our Siamese network \ref{chap:architecture}, how data is processed in Section~\ref{chap:preprocessing} and how we evaluate the performance of the channel chart in Section~\ref{chap:performance-metrics}.

\subsection{Local CIR Distance} \label{chap:cir_dist}

\noindent The idea of CC is that CSI measurements are spatially correlated as the CSIs are similar at the same position and become more and more dissimilar with their Euclidean distance in space. Hence, the goal of CC is to find a metric which (for any $i$ and $j$) resembles the Euclidean distance so that
\begin{equation}
    d_\mathsf{euc}(\ve{x}_i,\ve{x}_j) \propto d_\mathsf{csi}(\ve{\Tilde{h}}_i, \ve{\Tilde{h}}_j) \;,
\end{equation}
where $\ve{x}_i$ and $\ve{x}_j$ are two real-world coordinates and $\ve{\Tilde{h}}_i$ and $\ve{\Tilde{h}}_j$ their approximated channel impulse responses (CIRs) measured at the receiver. The channel model of a radio signal can be defined as
\begin{equation}
    h(t) = \sum_{n=0}^{N_\mathsf{p}-1} a_n \dirac(t - T_n) \;,
\end{equation}
where $N_\mathsf{p}$ is the total number of arriving signal paths from the transmitter at the receiver, $n$ is the index of the current path, $a_n$ is the complex gain of the $n$\textsuperscript{th} MPC, $\dirac(\cdot)$ is the Dirac delta function, and $T_n$ the delay of the component. To measure the CIR, a bandwidth-limited measurement signal $s(t)$ is first transmitted and 
\begin{equation}
    y(t) = h(t) \ast s(t) \;,
\end{equation}
defines the (noise-free) signal received at the base station, where $\ast$ is the convolution operator. The approximate CIR is then obtained from the autocorrelation: 
\begin{equation}
    \Tilde{h}(\tau) = \int_{-\infty}^\infty y(t)  s^\ast (t - \tau)\, \mathrm{d}t\;.
\end{equation}
The CIR of length $T$ for a measurement snapshot $i$ for base station $k$ is given in vector form as
\begin{equation}
    \vee{\Tilde{h}}_i^{(k)} \defeq \big[ \Tilde{h}^{(k)}_i(0),\, \ldots,\, \Tilde{h}^{(k)}_i(T-1) \big] \;.
\end{equation}

\begin{figure}[t]%
	\centering%
    \input{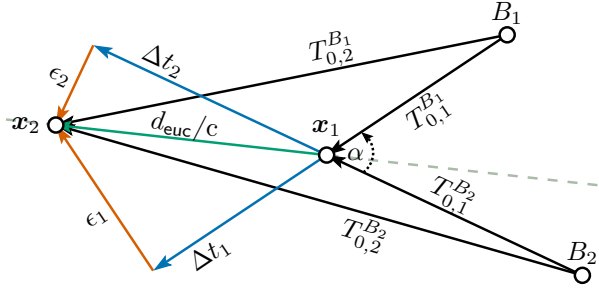}
	\caption{Schematic view of a radio environment with two base stations $B_1$ and $B_2$ showing the geometric context of ToF measurements w.r.t.\ the Euclidean distance of a displacement of a radio unit from $\ve{x}_1$ to $\ve{x}_2$. The above-shown quantities are the (time-)lengths of the described vectors.}
    \label{fig:dist_col}
\end{figure}%

In the following, we will base the derivation on the time of flight (ToF). For the time-difference-of-arrival (TDoA) case, the differences in the derivation can be found in Appendix~\ref{appendix:tdoa}. 
For two measurements that share the same MPCs in the same order, we define the ToF as 
\begin{equation}
    \Delta t_{n,i \cs j}^{} \defeq \big|T_{n,i}^{B_k} - T_{n,j}^{B_k} \big| \;,
\end{equation}
for the positions $\ve{x}_i$ and $\ve{x}_j$ of the CSI measurements $\ve{\Tilde{h}}_i$ and $\ve{\Tilde{h}}_j$ obtained at base station $B_k$. $T_{n,i}^{B_k}$ and $T_{n,j}^{B_k}$ are the time delays of the $n^{\text{th}}$ MPC from basestation $B_k$ at position $i$ and $j$. If the displacement of the position $\ve{x}_j$ is collinear to both the position $\ve{x}_i$ and the position of the base station $\ve{x}_b$, then
\begin{equation}
    d_\mathsf{euc}(\ve{x}_i, \ve{x}_j) = \mathrm{c} \, \Delta  t_{n,i \cs j} \;,
\end{equation}
where $\mathrm{c}$ is the speed of light. However, if they are not collinear the distance approximation is erroneous and can be defined as
\begin{equation} \label{eq:col_distance}
    d_\mathsf{euc}^{}(\ve{x}_i^{}, \ve{x}_j^{}) \defeq \mathrm{c} \sqrt{\left(\Delta t_{n,i \cs j}\right)^2 + \epsilon_{n,i \cs j}^2} \;,
\end{equation}
where the error $\epsilon_{n,i \cs j}$ is assumed to be orthogonal to the time difference $\Delta t_{n,i \cs j}$, as shown in Fig.~\ref{fig:dist_col}.\footnote{In the figure and derivations, $n=1$, $i=2$ and $j=1$ are assumed whenever not explicitly given, e.g., $\epsilon_{1,2\cs 1}^2 \equiv \epsilon_{1}^2$.}
Following the example case with two base stations shown in Fig.~\ref{fig:dist_col}, base station $B_1$ has a delay of the first direct path $T^{B_1}_{0,1}$ to position $x_1$ and a delay of $T^{B_1}_{0,2}$ to position $\ve{x}_2$. The vector difference of the delay $\Delta t_1$ to the expected delay $d_\mathsf{euc} / \mathrm{c}$ can be expressed as the error $\epsilon_1$:  
\begin{equation}
    \epsilon_1^2 = \frac{d_\mathsf{euc}^2}{ \mathrm{c}^2} - (\Delta t_1^{})^2_{} \;.
\end{equation}
As we do not know the direction of movement, we assume to have the similar likelihood for every direction, which leads to a arcsine distribution with a boundary of $[0, d_\mathsf{euc} / \mathrm{c}]$ for the error $\epsilon$. This means that the standard deviation of the error is monotonically increasing with the distance of the two measurements. If we have multiple base stations, e.g., $B_1$ and $B_2$, the errors $\epsilon_1$ and $\epsilon_2$ are geometrically related
\begin{equation}
    \epsilon_2 = \frac{d_\mathsf{euc}}{\mathrm{c}}
    \cos \left( \pi-\alpha \mp \cos^{-1} \bigs( \frac{\mathrm{c} \, \epsilon_1 }{d_\mathsf{euc}} \bigs) \right) \;,
\end{equation}
where the sign in front of the $\cos(\cdot)^{-1}$ depends on whether $B_1$ and $B_2$ lie on opposite sides of the line of motion of the UE (i.e., the green line between $x_1$ and $x_2$ in Fig.~\ref{fig:dist_col}) or on the same side. This means that for every $\alpha > 0$ the sum of the errors $\epsilon_1$ and $\epsilon_2$ have a lower bound on the the sum of distances
\begin{equation} \label{eq:euc_error}
    2 \, d_\mathsf{euc} > \mathrm{c} \, (\epsilon_1 + \epsilon_2) \;.
\end{equation}
This relation also holds for every MPC arriving at the receiver and allows us therefore to define a distance metric 
\begin{equation} \label{eq:distance_metric}
    d_\mathsf{s}^{}(\ve{\Tilde{h}}_i^{}, \ve{\Tilde{h}}_j^{}) \defeq \sum_{k=0}^{N_\mathsf{b}-1} \sum_{n=0}^{N_\mathsf{p}-1} \Delta t^{(k)}_{n,i \cs j} \;,
\end{equation}
which has a monotonically increasing mean and standard deviation, and a lower bound of the error. To calculate the distance metric, the estimation of the time delays of every arriving path is necessary. $\Delta t^{(k)}_{n,i \cs j}$ in \eqref{eq:distance_metric} denotes the time-difference observed at the $(k)$\textsuperscript{th} BS. However, as the extraction of the delay of the MPCs from measured CIRs is very challenging due to the bandwidth limited signal $s(t)$ \cite{kram2022delay}, we consider a simple approximation by subtracting the time-aligned CIRs:
\begin{equation} \label{eq:dist_approx}
    d^\prime_\mathsf{s}(\ve{\Tilde{h}}_i^{}, \ve{\Tilde{h}}_j^{}) \defeq \sum_{k=0}^{N_\mathsf{b}-1} \sum_{t=0}^{T-1} \big| |\Tilde{h}_i^{(k)}(t)| - |\Tilde{h}_j^{(k)}(t)| \big| \;,
\end{equation}
where $k$ is the index of the $N_\mathsf{b}$ base stations and $t$ is the time-index of the CIR of total length $T$. 

The limited window of every CIR includes the absolute position of the paths w.r.t.\ their ToF or TDoA, where the window is aligned by the first direct path of arrival (FDPoA). If the CIRs are recorded at the same position, the distribution of power within the CIRs is equal, which leads to ${d^\prime_\mathrm{s}(\ve{\Tilde{h}}_i^{}, \ve{\Tilde{h}}_i^{}) = 0}$, whereas measurements at different positions have a time shift of the paths within the CIR and therefore also a shift of the power. This leads to less overlap within the CIRs and therefore to a higher absolute difference, which is proportional to the sum of time differences of the arriving paths. As long as the main lobes of the bandwidth-limited signals are overlapping in the CIRs, the Euclidean distance and the CIR distance have a non-linear relationship, and have a linear one if the displacements between two CIR measurements is small. (The detailed derivation of this linear relationship can be found in Appendix~\ref{appendix:cir_linear}.) We therefore have
\begin{equation}
    {d_\mathsf{euc}^{}(\ve{x}_i^{}, \ve{x}_j^{}) \propto d_\mathsf{s}^{}(\ve{\Tilde{h}}_i^{}, \ve{\Tilde{h}}_j^{}) \propto d^\prime_\mathsf{s}(\ve{\Tilde{h}}_i^{}, \ve{\Tilde{h}}_j^{})} \;
\end{equation}
for small distances between CSI measurements, which means that our metric is locally restricted to its spatial neighborhood.


\subsection{Global Geodesic Distances with Siamese Networks} \label{chap:geo_dist}

\noindent The distance metric $d^\prime_\mathrm{s}(\ve{\Tilde{h}}_i^{}, \ve{\Tilde{h}}_j^{})$ between CSIs is restricted to short distances between two CSI measurements. Hence, our optimization is restricted to each sample's neighborhood. This only yields a local similarity and consistency of the channel chart. However, to use CC for localization we have to learn a channel chart reflecting the global geometry of the environment. To achieve this, we use the idea of Isomap~\cite{tenenbaum2000global} which creates global distances as the sum of local distances on the shortest paths on the manifold, i.e., \emph{geodesic distances}. Here, we estimate a matrix of pair-wise distances 
\begin{equation}
    \ve{D}_\mathsf{pw} \defeq 
    \begin{bmatrix}
         d^\prime_\mathrm{s}(\ve{\Tilde{h}}_{0}^{}, \ve{\Tilde{h}}_{0}^{}) & \cdots &  d^\prime_\mathrm{s}(\ve{\Tilde{h}}_{0}^{}, \ve{\Tilde{h}}_{N-1}^{})
         \\
         \vdots & \ddots & \vdots \\
         d^\prime_\mathrm{s}(\ve{\Tilde{h}}_{N-1}^{}, \ve{\Tilde{h}}_{0}^{}) & \cdots & d^\prime_\mathrm{s}(\ve{\Tilde{h}}_{N-1}^{}, \ve{\Tilde{h}}_{N-1}^{})
    \end{bmatrix} \;,
\end{equation}
$\ve{D}_\mathsf{pw} \in \mathbb{R}^{N\times N}$, to create a neighborhood graph. The shortest paths between each coordinates of the channel chart $d^\prime_\mathrm{s}(\ve{\Tilde{h}}_i^{},\ve{\Tilde{h}}_j^{})$ are then estimated using a shortest-path estimator, e.g., via the Dijkstra algorithm \cite{dijkstra1959note}.\footnote{Note that we do not need to run the estimator for each of the pairs in the matrix separately as due to the \emph{principle of optimality}~\cite{Kirk1970}, any pairs of points on a sub-path of a shortest path themselves constitute an optimal path.} This allows us to define valid distances for all points, also for far points, on the manifold 
\begin{equation}
    \ve{D}_\mathsf{geo} \defeq 
    \begin{bmatrix}
         d^\prime_\mathrm{geo}(\ve{\Tilde{h}}_{0}^{}, \ve{\Tilde{h}}_{0}^{}) & \cdots &  d^\prime_\mathrm{geo}(\ve{\Tilde{h}}_{0}^{}, \ve{\Tilde{h}}_{N-1}^{})
         \\
         \vdots & \ddots & \vdots \\
         d^\prime_\mathrm{geo}(\ve{\Tilde{h}}_{N-1}^{}, \ve{\Tilde{h}}_{0}^{}) & \cdots & d^\prime_\mathrm{geo}(\ve{\Tilde{h}}_{N-1}^{}, \ve{\Tilde{h}}_{N-1}^{})
    \end{bmatrix} \;,
\end{equation}
$\ve{D}_\mathsf{geo} \in \mathbb{R}^{N\times N}$, by summing up the local distances on the shortest path between two points 
\begin{equation}
    d_\mathsf{geo}^\prime(\ve{\Tilde{h}}_i^{},\ve{\Tilde{h}}_j^{}) \defeq \sum_{p \in \mathcal{P}} d^\prime_{\mathrm{s}}(\ve{\Tilde{h}}_{p}^{}, \ve{\Tilde{h}}_{p+1}^{}) 
      \;, 
\end{equation}
where $p$ and $p+1$ are the indices of all neighboring pairs of points on the shortest path $\mathcal{P}$. As our CIR distance is linear with a constant slope for small displacements, we also have a linear geodesic distance if the spatial density of neighbors is high, as the sum of linear elements with the same slope will also be linear. 

Intentionally, Isomap performs dimensionality reduction on an enclosed data set, which therefore restricts the method to predict the embedding on unseen data. The full geodesic distance matrix has to be calculated including the previous and new data to predict on unseen data, which is both time-consuming and impractical for real-time locating systems. 
Hence, at first glance, Isomap is inappropriate for CC, that predicts on unseen data. While there are extensions of the original Isomap algorithm, such as the landmark Isomap \cite{de2004sparse}, which can interpolate the embedding on unseen data, Dimal et al.~\cite{pai2019dimal} have shown that Siamese networks outperform the classical approaches in terms of generalization. In this approach the multidimensional scaling (MDS) is replaced by a Siamese network \cite{chicco2021siamese}, which encodes the input data into the 2D space and then learns the distance between two inputs in the embedding given a distance metric. The goal of the Siamese network is to learn the proposed geodesic distance $d_\mathsf{geo}^\prime(\ve{\Tilde{h}}_i^{},\ve{\Tilde{h}}_j^{})$ with the objective
\begin{equation}
    \mathcal{L} = \big|d_\mathsf{geo}^\prime(\ve{\Tilde{h}}_i^{},\ve{\Tilde{h}}_j^{})-\|\ve{z}_i^{}-\ve{z}_j^{}\|_2^{} \big| \;,
\end{equation}
where $\ve{z}_i$ and $\ve{z}_j$ are the two-dimensional outputs of the neural network for the inputs $\ve{\Tilde{h}}_i^{}$ and $\ve{\Tilde{h}}_j^{}$. Instead of utilizing all data like in the Isomap algorithm, we sample randomly from the available geodesic data until the Siamese network converged. After optimization, the neural network has learned the global geometry of the radio environment. As the neural network uses the CSI measurements as input data, the network learns a transformation from the manifold of the input data to a 2D embedding. This allows the network to interpolate and therefore also to efficiently process unseen data without recalculating the geodesic distance matrix.

\subsection{Architecture} \label{chap:architecture}

\noindent As CC is very similar to fingerprinting, we employ a similar efficient architecture as proposed in \cite{stahlke20225g}. The network consists of 4 convolutional layers and 2 dense layers. Batch normalization stabilizes the training and we apply Rectified Linear Units (ReLU) as activation functions except for the last layer, where we applied no activation function. We have no local pooling layers between the convolutional layers to keep the dimension of time, which has shown good results in time series downstream tasks \cite{ismail2019deep}. Instead, we use a global average pooling of the channels before the dense layers. The kernel sizes of the convolutional layers are increased with the depth of the model to enhance the receptive field of the network. The kernel sizes being used depends on the effective input resolution of the CIRs. As we have two different radio systems with different resolutions of the CIRs, we have different kernel sizes of the neural network adapted for the respective bandwidth: (i) for the UWB data we have kernel sizes of ${[3 \times 3]}$ for the first, ${[5 \times 5]}$ for the second, ${[15 \times 15]}$ for the third, and ${[30 \times 30]}$ for the last layer, and (ii) for the 5G data we have ${[3 \times 3]}$ for the first, ${[5 \times 5]}$ for the second, ${[8 \times 8]}$ for the third, and ${[10 \times 10]}$ for the last layer. The last layers estimates the embedding, which is our 2D embedding of the input CSI data.

\subsection{Preprocessing}\label{chap:preprocessing}

\noindent For data preprocessing, we follow the idea from \cite{stahlke20225g} to exploit both the raw CIR in the time domain and the corresponding ToF/TDoA using a CNN. We generate a 2D tensor of dimensions $[N_\mathsf{A},L_\mathsf{w}]$, with $N_\mathsf{A}$ anchors and $L_\mathsf{w}$ time-steps in the temporal resolution of the CIR. The CIRs are padded by the corresponding ToF/TDoA within the tensor to model the unique relative shift of the FDPoA for every position in the area.

\subsection{Performance Metrics}\label{chap:performance-metrics}
To measure the performance of a generated channel chart, we apply continuity (CT) and trustworthiness (TW) to measure its local similarity \cite{studer2018channel}. CT measures whether a chart introduces wrong nearest neighbors in the embedding and penalizes them proportional to the rank of the nearest neighbors 
\begin{equation}
    \mathrm{CT}(K) =1-\dfrac{2}{K(2N-3K-1)}\sum\limits_i^N\sum\limits_{{j \in \mathcal{V}_{K}(v_{i})}}\big(\hat{r}(i,j)-K \big)  \;.
\end{equation}
$\mathcal{V}_{K}(v_{i})$ are the $K$ nearest neighbors in the original space, $N$ is the total number of samples and $\hat{r}(i, j)$ is the rank among the pair-wise distances in the embedding. Conversely, TW measures whether the original space has different nearest neighbors as modelled in the channel chart
\begin{equation}
    \mathrm{TW}(K) =1-\dfrac{2}{K(2N-3K-1)}\sum\limits_i^N\sum\limits_{{j \in \mathcal{U}_{K}(u_{i})}}\big(r(i,j)-K \big) \;.
\end{equation}
Here, $\mathcal{U}_{K}(u_{i})$ are the $K$ nearest neighbors in the chart and $r(i, j)$ is the rank among the pair-wise distances in the original space. 

\begin{figure}[t]%
	\centering%
	\begin{subfigure}{.455\columnwidth}%
		\includegraphics[clip, width=\columnwidth]{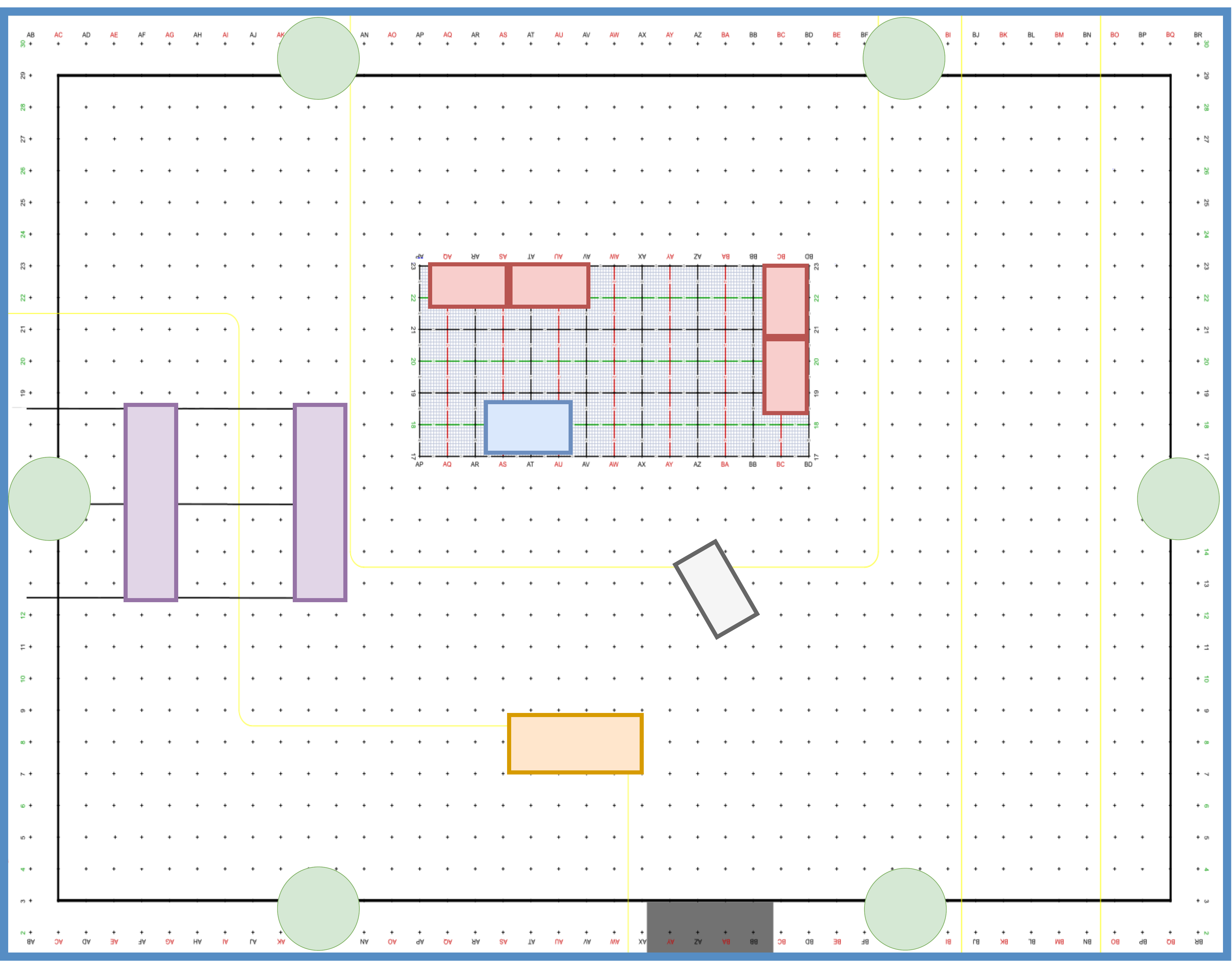}%
	\end{subfigure}%
	\hspace{2pt}%
	\begin{subfigure}{.525\columnwidth}%
		\includegraphics[clip, width=\columnwidth]{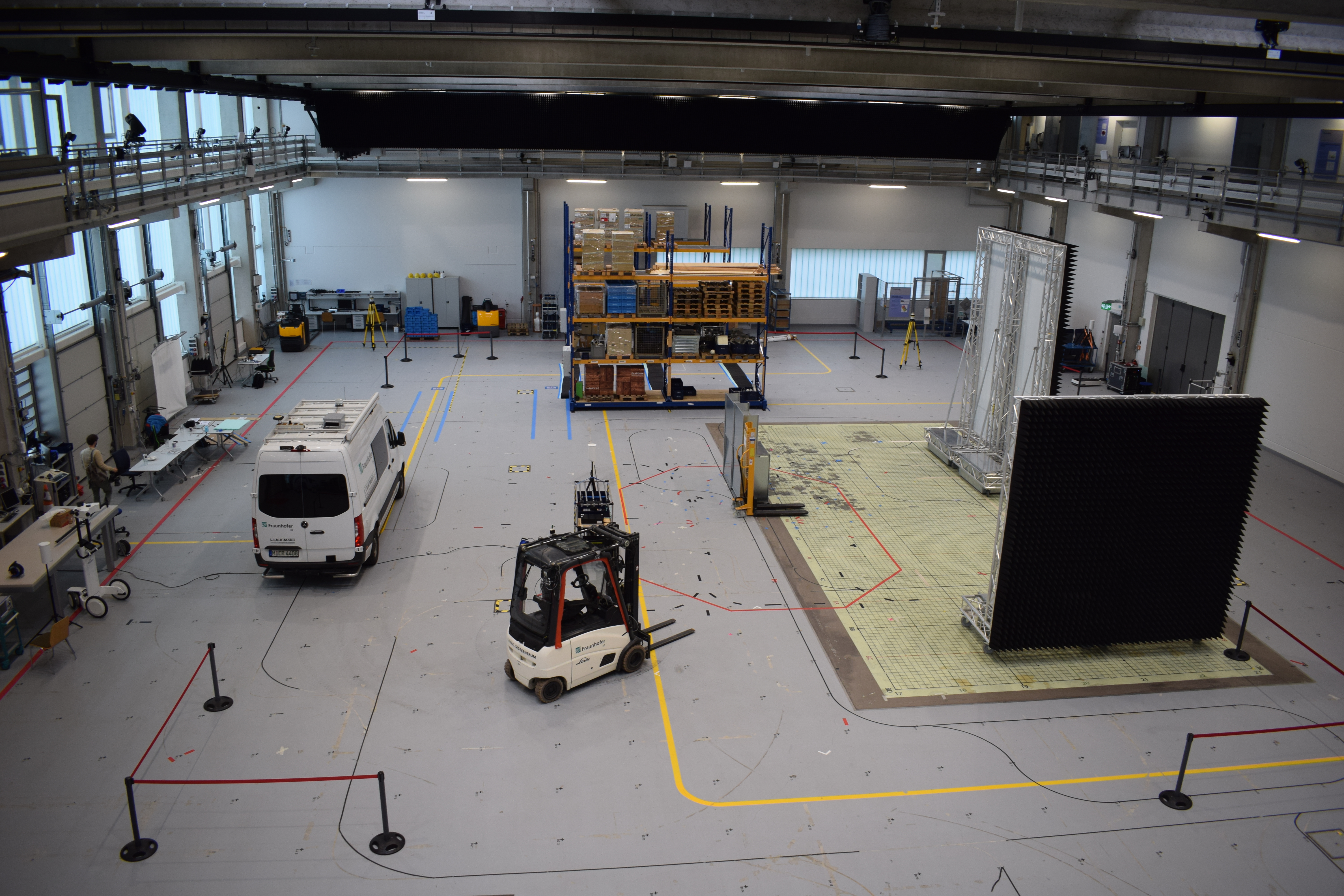}%
	\end{subfigure}%
	\caption{Schematic top view (left) of the environment (right). The environment consists of reflective walls (red), an industrial truck (orange), a fork lift (gray), large shelves (purple), and a small shelf (blue). The BSs are indicated as green dots and are placed at a height of 6-7\,m. 
	}%
	\label{fig:environment_5g}
\end{figure}%
As proposed in \cite{studer2018channel}, we use $K = 0.05N$ as the number of nearest neighbors to calculate TW and CT. Both metrics have a value range of $[0,1]$, where $1$ means that the local geometry is perfectly preserved and $0$ that the local geometry is not preserved.

However, to measure the capabilities of CC for positioning, we have to measure if the channel chart reflects the global geometry. As we do not train our channel chart using side information, we follow the idea of Pihlajasalo et al.~\cite{pihlajasalo2020absolute} to estimate an affine transformation to transform the local coordinate frame of the channel chart into the global coordinate frame, which does not affect its local structure as it is only a linear transformation. Theoretically, we only need 3 points to define an affine transformation between two coordinate systems in the 2D space. However, as a chart will always have imperfections, we cannot select the optimal points in it to do the transformation. Instead, we use all points to estimate a transformation matrix with a least-squares solution 
\begin{equation}
   \ve{\hat{A}} \defeq \argmin_{\ve{A}} \|\ve{A} \ve{z}- \ve{x}\|_{2}^{2} \;,
\end{equation}
where $\ve{\hat{A}}$ is the estimated transformation matrix, $\ve{z}$ are the points of the channel chart, and $\ve{x}$ are the coordinates in the real-world frame. We calculate the mean absolute error (MAE), with the error being the Euclidean distance between the channel-chart points and ground truth, and the 90\textsuperscript{th} percentile of the cumulative distribution function (CDF) of the error (CE90) as performance metric.

\section{Experimental Setup}
\label{section:experimental_setup}

\subsection{Data}

\noindent To show the capabilities of CC for positioning, we compare the performance with two different radio setups. The first setup is a 5G downlink time-difference-of-arrival (TDoA) setup and the second setup is a UWB time-of-flight (ToF) radio setup.

\textbf{5G}. For our first experiment, we use a 5G downlink TDoA setup with six commercial off-the-shelf software-defined-radio BS. The radio system has a bandwidth of 100\,MHz and a center frequency of 3.7\,GHz. The BS transmit power is set to 20\,dBm. All BSs are highly synchronized by means of a common signal generator. The recording frequency is 6.6\,Hz. Fig.~\ref{fig:environment_5g} shows a schematic view and the real-world environment. The transceivers (green) are placed at the edges of the recording area at a height of 6-7\,m. We created a typical industrial setup with reflective walls (red), an industrial truck (orange), a forklift (gray), and small (blue) and large shelves (purple) to block the LoS to the receiver and to create dense multipath propagation. The receiver is placed at a height of 1.95\,m on a handcart, which is moved by a person. Also for the 5G setup, we record a training data set with 18,722 bursts and a test data set with 15,721 bursts on different trajectories within the same environment.

\begin{figure}[t]%
	\centering%
	\begin{subfigure}{.455\columnwidth}%
		\includegraphics[clip, width=\columnwidth]{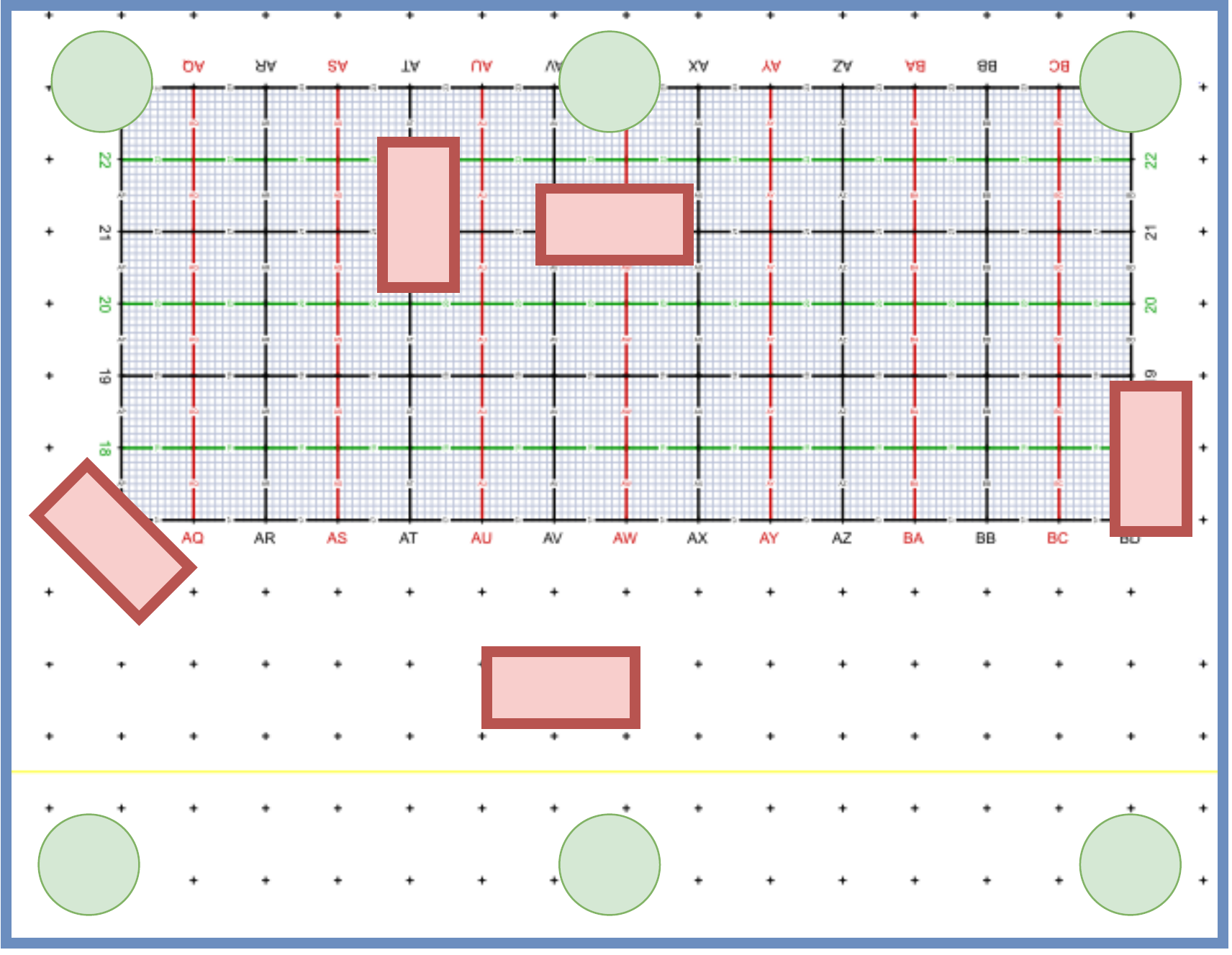}%
	\end{subfigure}%
	\hspace{2pt}%
	\begin{subfigure}{.525\columnwidth}%
		\includegraphics[clip, width=\columnwidth]{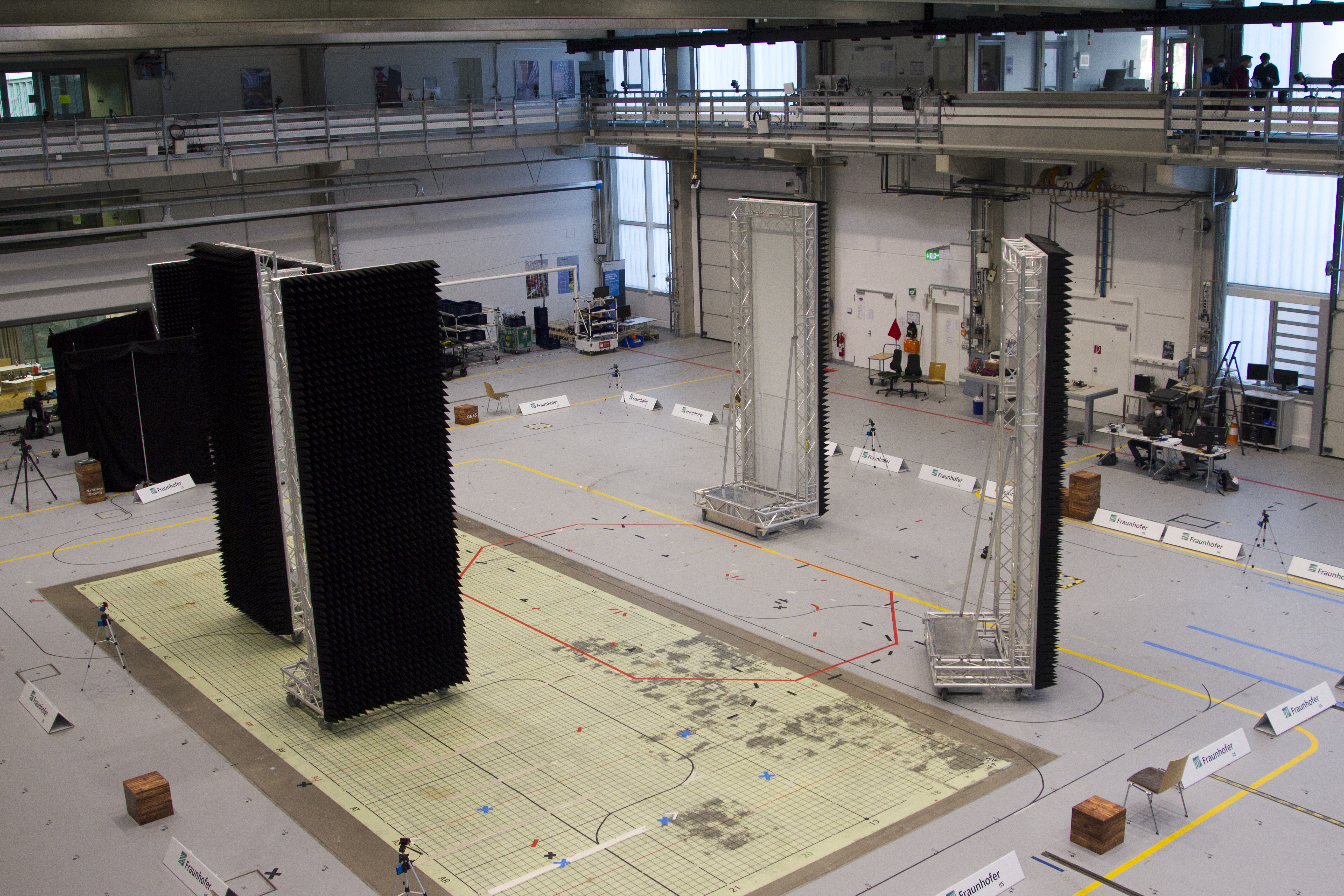}%
	\end{subfigure}%
	\caption{Schematic top view (left) of the environment (right). The red rectangles indicate reflective walls and the green dots are the stationary UWB transceiver modules.}%
	\label{fig:environment_uwb}
\end{figure}%

\textbf{UWB}. For our second experiment, we use a UWB radio setup with 6 transceivers acting as base station (BS) and one recording robot platform with a maximum velocity of $v_\mathsf{max}=0.2\,\mathrm{m} \mathrm{s}^{-1}$. We configured the system for ToF acquisition as a two-way ranging setup with a bandwidth of 499.2\,MHz at a center frequency of 4\,GHz with a recording frequency of 3\,Hz. We designed a complex environment with walls, that reflect radio signals on the inner side (iron surface) and absorb them at the outside (black surface). Fig.~\ref{fig:environment_uwb} shows a schematic view and the real-world environment. The transceivers, indicated as green dots, are placed at the edges of the recording area, shown as blue rectangles. The reflective walls, indicated in red, are placed to block the LoS between the anchors and the robot platform, which causes ranging errors of the UWB radio system leading to high localization errors using classical positioning approaches. We recorded two different data sets: a training data set with 18.027 bursts, where one burst includes the 6 synchronized CIRs and ToF measurements, and a test data set with 3,382 bursts. Both data sets are recorded independently on different trajectories within the same area.

\subsection{Baselines}

\noindent In the following, we describe  all relevant state-of-the-art CC methods that we compare our solution with.

\subsubsection{Non-parametric Approaches}

\noindent We compare our approach with PCA, Sammon mapping \cite{studer2018channel}, and Isomap \cite{le2021efficient}. For the pair-wise distances used in Sammon mapping, we applied the CIR distance defined in Section~\ref{chap:cir_dist}. For Isomap, we used the geodesic distance defined in Section~\ref{chap:geo_dist}.

\begin{figure}[t]%
	\centering%
	\begin{subfigure}{1\columnwidth}%
		\includegraphics[clip, width=\columnwidth]{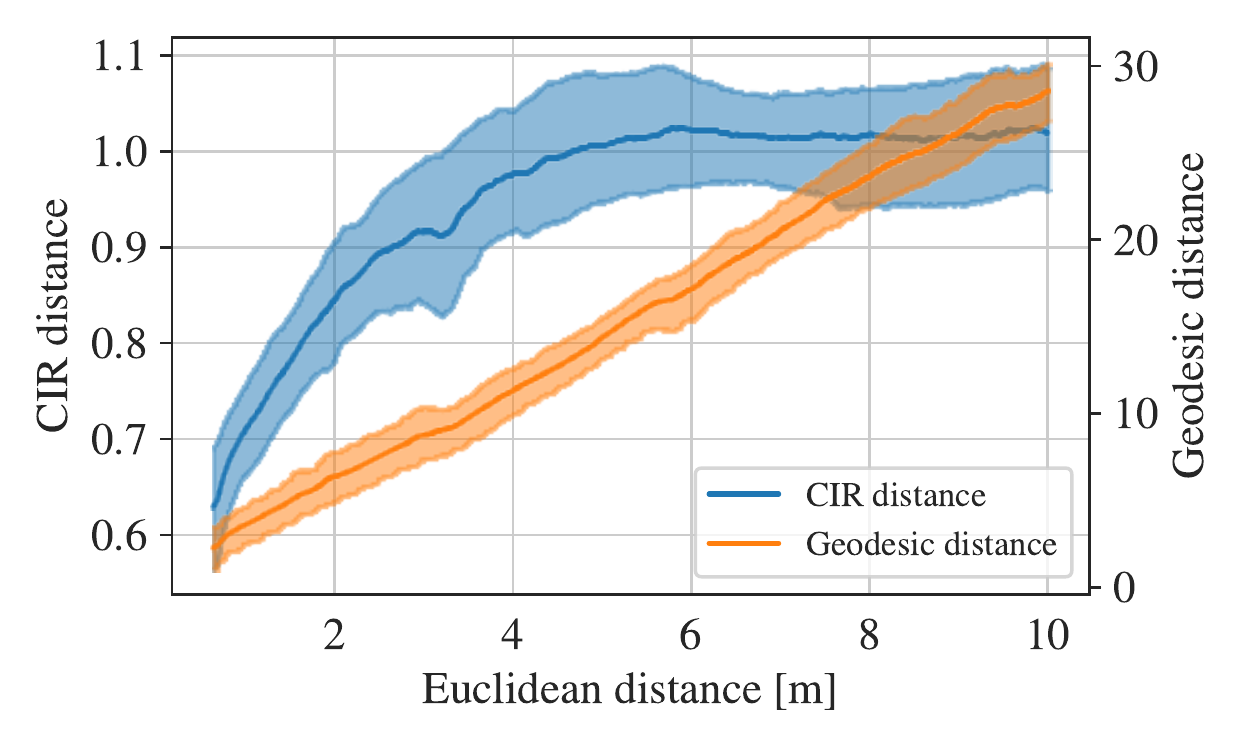}%
	\end{subfigure}%
	\caption{The proposed local CIR distance (blue) and the global geodesic distance (orange) for random combinations of positions in the UWB environment.}
    \label{fig:cir_geo_dist}
\end{figure}%
\subsubsection{Parametric Approaches}

In general, we investigate three different parametric baseline methods, as follows.

First, we apply the two stage approach from Agostini et al.~\cite{agostini2022not}. They used a convolutional autoencoder to compress the CSI to the quintessential information and used UMAP to create the channel chart. As autoencoder architecture, we use two convolutional layers with kernel size 3 for the first and 5 for the second layer. Both layers have 8 channels. As the decoder we use a fully connected layer to restore the input size with 2 consecutive transposed convolutional layers with kernel size 3 and 5 and also 8 channels. For all layers except the last layer, we used ReLU activation functions and for the convolutional layers batch normalization. The mean squared error (MSE) is used to evaluate the reconstruction error. We compressed the input data to 5\,\% of its size and applied UMAP on the latent variables with the Euclidean distance as metric. 

The second approach is the constrained autoencoder proposed by Huang et al.~\cite{huang2019improving}. They used a fully connected autoencoder to compress the input to the size of two, which are the coordinates of the channel chart. Besides the reconstruction error (MSE), they added a constraint which minimizes the pair-wise distances between two compressed input instances. As pair-wise distances, we use the CIR distance defined in Section~\ref{chap:cir_dist} between the raw input vectors. The architecture consists of 4 fully connected layers with [500, 100, 50, 20] neurons, while the encoder is the mirrored structure.

The third approach is proposed by Ferrand et al.~\cite{ferrand2021triplet} and uses triplets. The triplets are built using the recording time of the signals assuming that CSIs close in time are also close in space, due to the physical constraints of the movement, whereas samples that are farther in time are more distant in space. Hence, for the triplets, samples that are close in time (i.e., in our case within a window of $\pm$3\,s) are assigned as positive, and negative samples are assigned within a window of $\pm$2,400\,s but not within the positive window. These parameters are estimated experimentally and depend on the movement pattern. As architecture we use the same model as for our geodesic Siamese network defined in Section~\ref{chap:architecture}.



\section{Evaluation}
\label{section:evaluation}

\begin{figure}[t]%
	\centering%
	\begin{subfigure}{1\columnwidth}%
		\includegraphics[clip, width=\columnwidth]{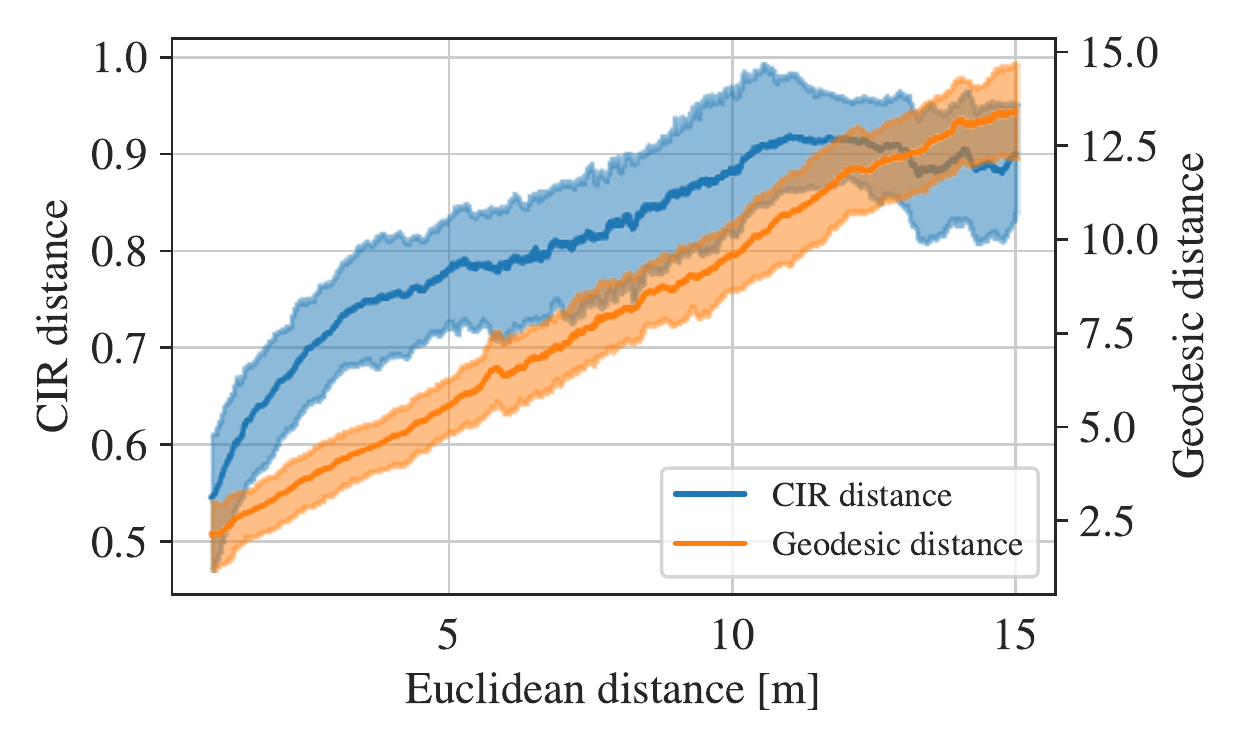}%
	\end{subfigure}%
	\caption{The proposed local CIR distance (blue) and the global geodesic distance (orange) for random combinations of positions in the 5G environment.}
    \label{fig:cir_geo_dist_5G}
\end{figure}%

\noindent We first evaluate the proposed global geodesic distance metric. Next, we compare the geodesic Siamese network with the relevant state-of-the-art of CC for different radio setups w.r.t.\ positioning. The CC algorithms are trained only on the training data set and are tested on both the training and test data to show their generalization capabilities on unseen data.

\subsection{Distance Metric Evaluation}

\noindent For the evaluation of the proposed geodesic distance metric described in Section~\ref{chap:geo_dist}, we use the UWB and 5G data sets, where we randomly sample measurements and calculate the distance metrics and the corresponding Euclidean distance in space. Fig.~\ref{fig:cir_geo_dist} shows the local CIR distance (blue) and the global geodesic distance (orange) for the UWB data set, where the Euclidean distance between two positions is shown on the $x$-axis and the proposed distances are shown on the $y$-axis. As described in Section~\ref{chap:cir_dist}, we can clearly see that the CIR distance is only (linearly) correlated to the Euclidean distance for points that are close in space. From the CIR distance we can neither reliably (as the relation is constantly changing) nor unambiguously (as the CIR distance is not invertible) derive an Euclidean distance for two points that are farther apart.

The mean and the standard deviation of the error increases monotonically with the Euclidean distance until about 6\,m. For Euclidean distances larger then 6\,m, the CIR distance has no more correlations to the Euclidean distance, which restricts the measurements to their neighbors. We can see a similar behaviour for the 5G data set shown in Fig.~\ref{fig:cir_geo_dist_5G}. The CIR distance is correlated to the Euclidean distance until about 10\,m. Due to the lower bandwidth of the 5G signals, the lobe of the received signals are wider, which leads correlations of signals for higher Euclidean distances compared to signals with lower bandwidths. However, this also leads to a higher noise level due to power variations in the CIR because of fast fading effects, which is a violation of our distance metric model. To have a correlation of the radio signals also for larger distances, we propose to create geodesic distances from a linear combination of local CIR distances as described in Section~\ref{chap:geo_dist}.

For both radio systems, the geodesic distance shows a high linear correlation to the Euclidean distance, shown in Fig.~\ref{fig:cir_geo_dist} and Fig.~\ref{fig:cir_geo_dist_5G} in orange. This means that the spatial density of CSI measurements is high enough to provide a good linear approximation of the CIR distance and therefore also of the geodesic distance. In contrast to the CIR distance, the geodesic distance is globally valid, which means that the metric correlates to the Euclidean distance for arbitrary large distances. This allows us to optimize channel charts with a global similarity, which is necessary for CC-assisted localization.

\begin{figure*}[t]
    \centering
    \subfloat[Radio environment\label{fig:5g_re}]{
        \includegraphics[trim={5mm 2mm 0 0}, width=.32\linewidth]{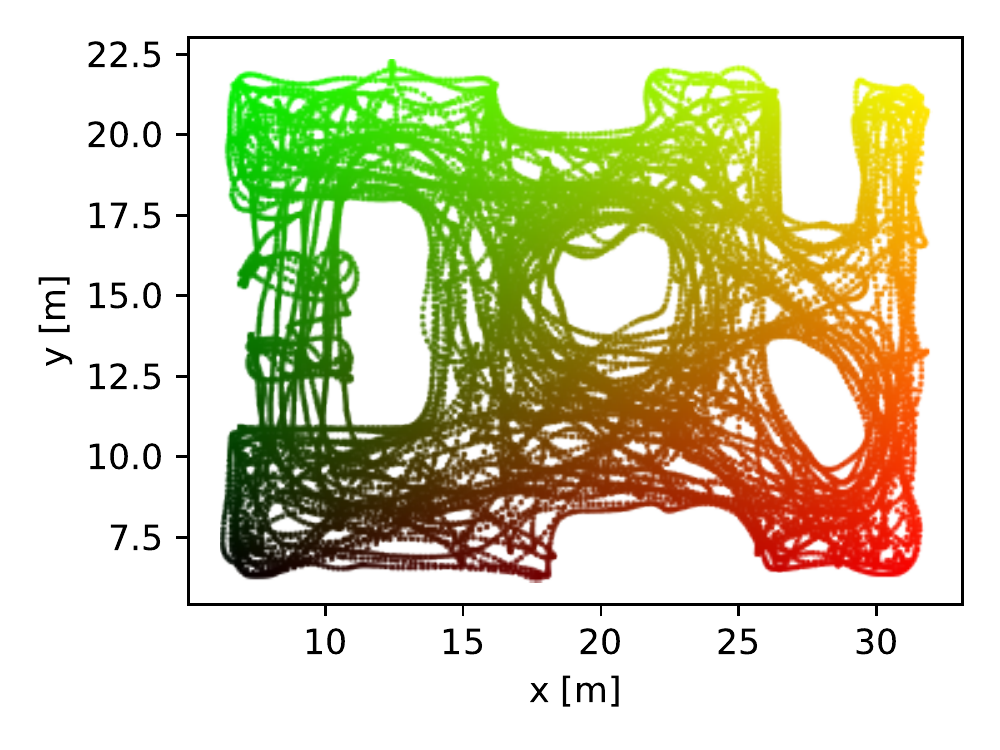}}
    \subfloat[Siamese Geo.\label{fig:5g_sg}]{
        \includegraphics[trim={5mm 2mm 0 0},width=.32\linewidth]{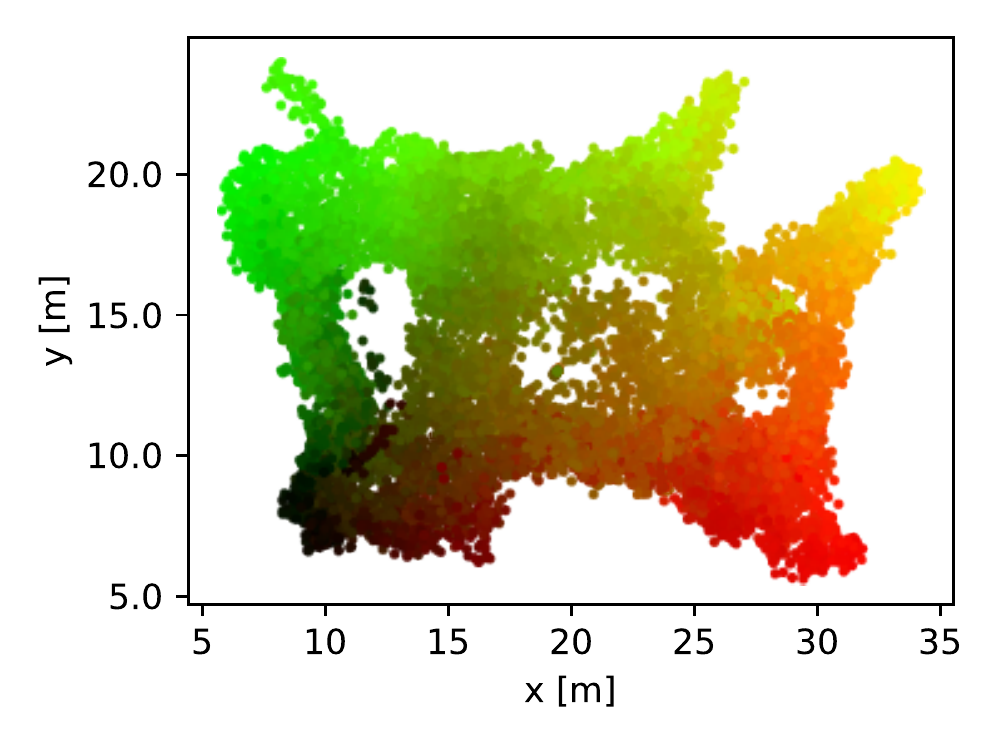}}
    \subfloat[CAE + UMAP \label{fig:5g_umap}]{
        \includegraphics[trim={5mm 2mm 0 0},width=.32\linewidth]{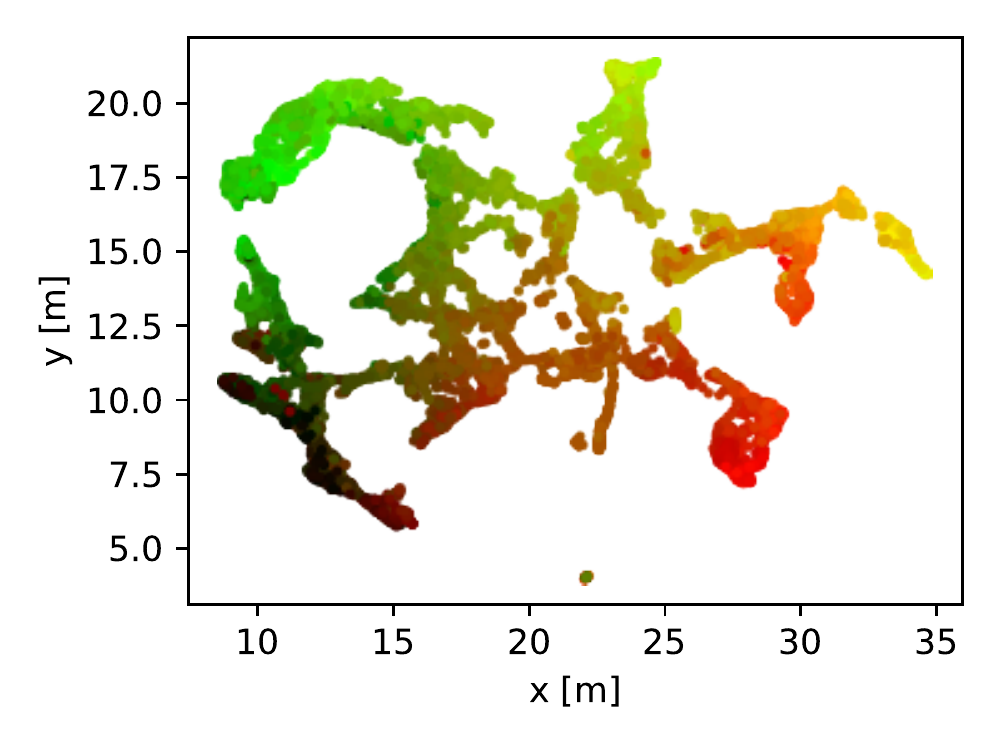}}
    \\
    \subfloat[Triplet\label{fig:5g_t}]{
        \includegraphics[trim={5mm 2mm 0 0},width=.32\linewidth]{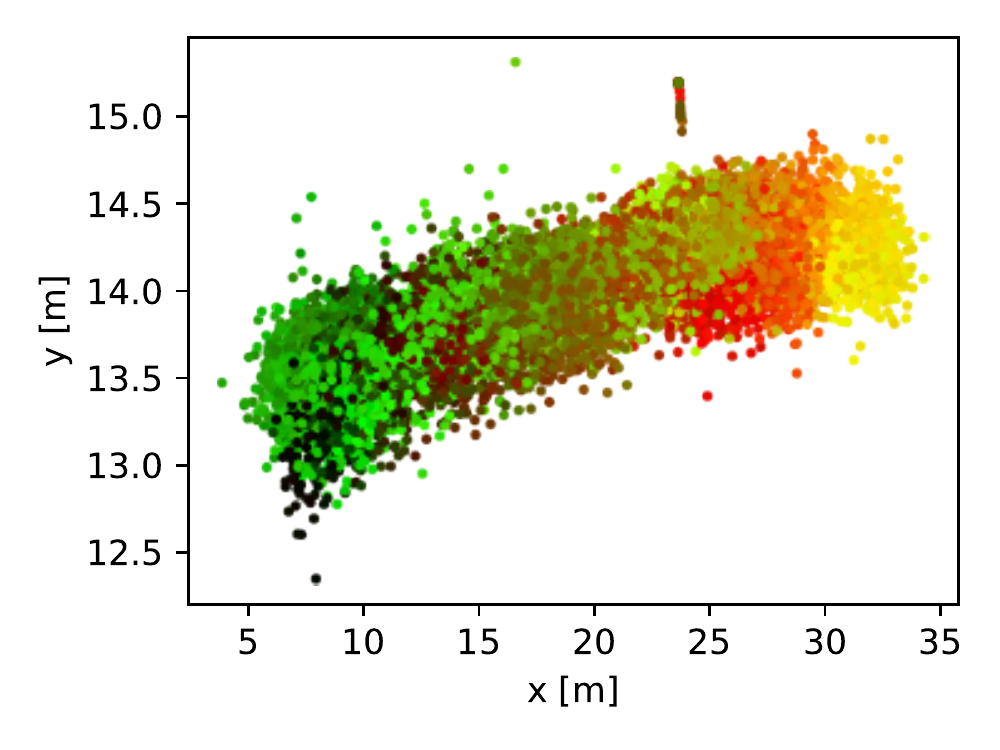}}
    \subfloat[Isomap\label{fig:5g_iso}]{
        \includegraphics[trim={5mm 2mm 0 0},width=.32\linewidth]{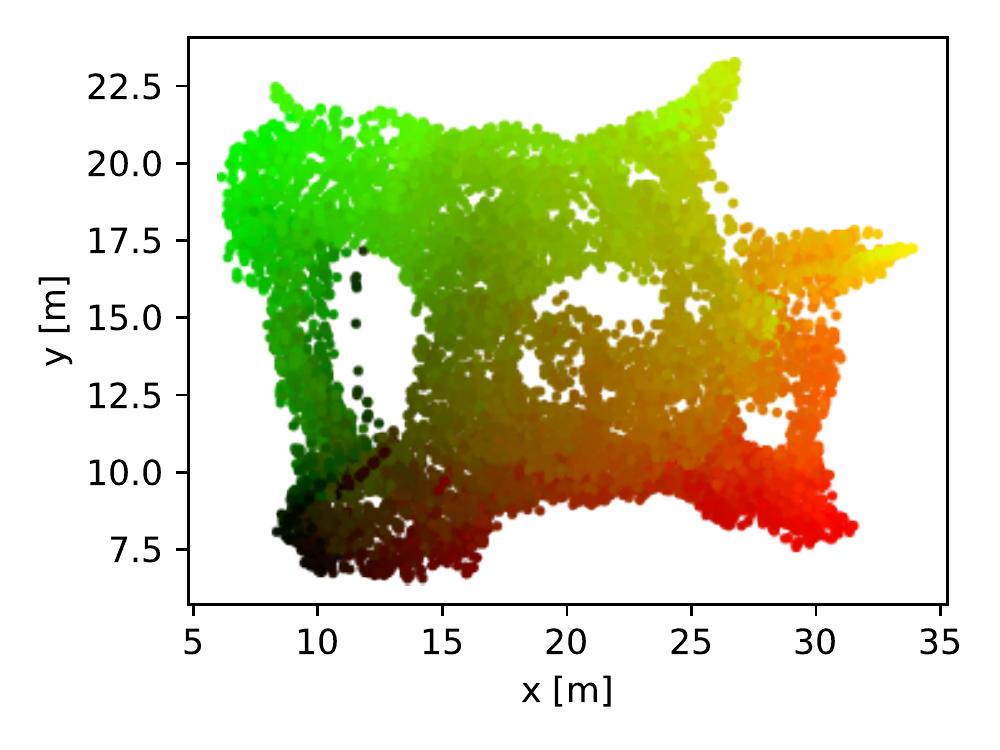}}
    \subfloat[Con.\ AE\label{fig:5g_cond}]{
        \includegraphics[trim={5mm 2mm 0 0},width=.32\linewidth]{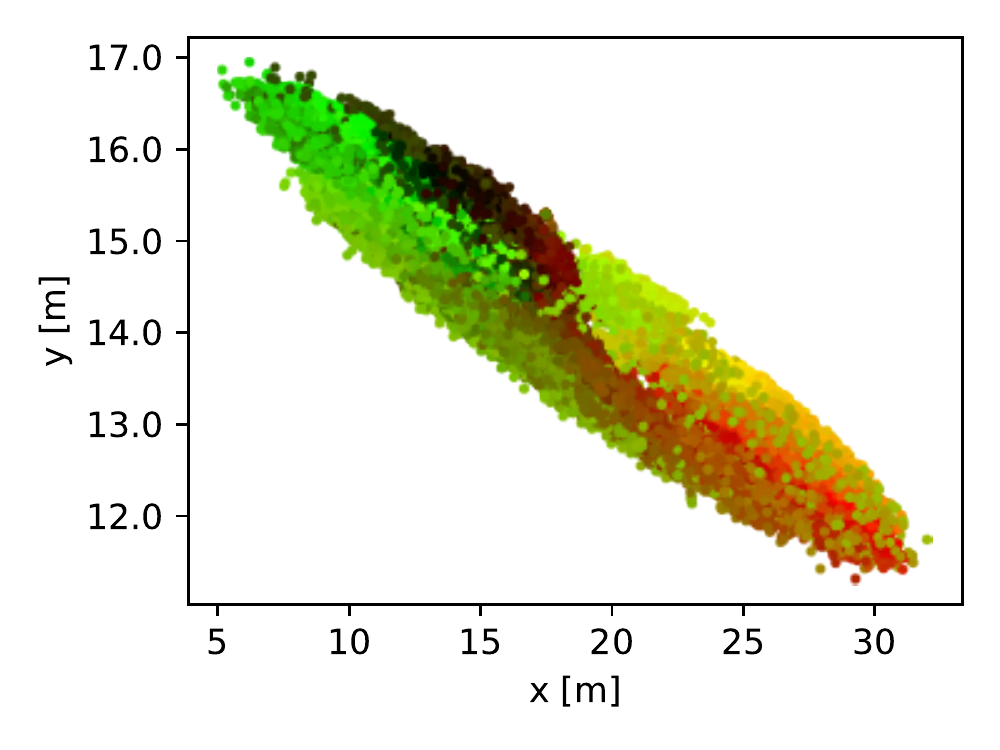}}
    \caption{Results CC methods on the 5G data set. Graph (a) shows the recorded training data in the real-world environment. The color gradient is used to to show the relation of the positions of the channel charts to the real positions. Graphs (b)--(f) show the channel charts after the affine transformation to the global coordinate frame. }
    \label{fig:results_5G}
\end{figure*}

\subsection{5G}

\begin{table}[t]
\caption{Results of CC methods on the 5G data set, tested on both the training and test data sets. The unit of CE90 and MAE is meter. }
\setlength{\tabcolsep}{3pt}
\resizebox{\columnwidth}{!}{
\begin{tabular}{lcccc|cccl}
\toprule
& \multicolumn{4}{c}{\textbf{Training}} & \multicolumn{4}{c}{\textbf{Test}}
\\\cmidrule(lr){2-5}\cmidrule(lr){6-9}
        \textbf{Method} & CT & TW & CE90 & MAE  & CT  & TW & CE90 & MAE\\\midrule
        Ours \ & 0.986 & 0.986 & 2.35 & 1.40 & 0.983 & 0.982 & 2.48 & 1.46 \\
        Isomap \cite{le2021efficient} & 0.985 & 0.986 & 2.46 & 1.42 & 0.984 & 0.984 & 2.58 & 1.47 \\
        CAE + UMAP \cite{agostini2022not} & 0.966 & 0.961 & 3.87 & 2.12 & 0.958 & 0.947 & 4.55 & 2.37\\
        Triplet \cite{ferrand2021triplet} & 0.920 & 0.824 & 6.90 & 4.17 & 0.919 & 0.842 & 6.69 & 3.89\\
        Con.\ AE \cite{huang2019improving} & 0.919 & 0.895 & 8.19 & 4.66 & 0.919 & 0.895 & 7.71 & 4.37 \\
        Sammon \cite{studer2018channel} & 0.899 & 0.880 & 7.41 & 4.04 & 0.897 & 0.890 & 6.68 & 3.77 \\
        PCA \cite{studer2018channel} & 0.918 & 0.860 & 7.59 & 4.16& 0.921 & 0.869 & 7.27 & 4.02\\
        \midrule
\end{tabular}}
\label{tab:tab_results_5G}
\end{table}

\noindent Table~\ref{tab:tab_results_5G} lists the results of CC methods on the 5G setup, where \emph{training} means that the models are evaluated on training data set and \emph{test} that the models are evaluated on the test data set. Our method shows along with Isomap the highest local similarities of $\mathrm{TW} > 0.98$, $\mathrm{CT} > 0.98$ and a global similarity of $\mathrm{CE90} < 2.46\,\mathrm{m}$, $\mathrm{MAE} < 1.42\,\mathrm{m}$, while our approach achieves a slightly better global accuracy. The main difference of our approach to the Isomap is mainly the runtime improvement, as Isomap has to recalculate the geodesic distance matrix for every unseen data points again and our approaches learned a transformation function enabling efficient prediction on unseen data. CAE + UMAP still achieves reasonable results of $\mathrm{TW}$ and $\mathrm{CT} = 0.96$ for the local similarity, while the global similarity is low with a $\mathrm{CE90} = 3.87\,\mathrm{m}$ and $\mathrm{MAE} = 2.12\,\mathrm{m}$. We also tested UMAP on the raw input embeddings with the CIR distance as pair-wise distance and achieved $\mathrm{CT} = 0.947$ and $\mathrm{TW} = 0.972$ for the local similarity and $\mathrm{CE90} = 8.74\,\mathrm{m}$ and $\mathrm{MAE} = 5.26\,\mathrm{m}$ for the global similarity. However as already investigated in \cite{agostini2022not}, we could not achieve higher results especially for the global similarity. The worst performance is achieved by PCA, Sammon mapping and the triplets approach with a local similarity $\mathrm{TW} < 0.89$ and $\mathrm{CT} < 0.93$ and a global similarity of $\mathrm{CE90} > 6.90\,\mathrm{m}$ and $\mathrm{MAE} > 4.00\,\mathrm{m}$. The PCA only learns a linear transformation and Sammon mapping considers the distances between all points, including the points that are far away. However, as distances between two CSI measurements are only valid in the proximity of two points, considering also far points introduces errors in the CC generation. The worse performance of the triplet approach can be explained by the assumption that far points in time are also far in space. This assumption depends highly on the movement pattern, and is therefore error prone, especially in the 5G data set as the receiver is moved by a person. The constrained autoencoder (Con.~AE) faces the same problem as Sammon mapping, as also distances from far points are considered during optimization. However, we think due to the unique mapping from the input space into the embedding, the autoencoder identifies wrong distance labels implicitly and mitigates the effect of wrong distance approximations, which leads to slightly better results as in Sammon mapping.

\begin{table}[t]
\caption{Results of CC methods on the UWB data set, tested on both the training and test data sets. The unit of CE90 and MAE is meter. }
\setlength{\tabcolsep}{3pt}
\resizebox{\columnwidth}{!}{
\begin{tabular}{lcccc|cccc}
\toprule
& \multicolumn{4}{c}{\textbf{Training}} & \multicolumn{4}{c}{\textbf{Test}}
\\\cmidrule(lr){2-5}\cmidrule(lr){6-9}
        \textbf{Method} & CT  & TW & CE90 & MAE & CT  & TW & CE90 & MAE\\\midrule
        Ours\ & 0.997  & 0.997 & 1.30 & 0.69 & 0.997 & 0.996 & 1.28 & 0.72 \\
        Isomap \cite{le2021efficient} & 0.997 & 0.997 & 1.33 & 0.72 & 0.996 & 0.996 & 1.34 & 0.80 \\
        CAE + UMAP \cite{agostini2022not} & 0.996 & 0.996 & 1.37 & 0.72 & 0.995 & 0.994 & 1.24 & 0.68\\
        Triplet \cite{ferrand2021triplet} & 0.975 & 0.951 & 3.98 & 2.29 & 0.973 & 0.948 & 3.94 & 2.25\\
        Con.\ AE \cite{huang2019improving} & 0.965 & 0.928 & 4.76 & 2.55 & 0.966 & 0.939 & 4.38 & 2.30 \\
        Sammon \cite{studer2018channel}  & 0.943 & 0.880 & 5.49 & 2.98 & 0.943 & 0.888 & 5.49 & 2.98 \\
        PCA \cite{studer2018channel}  & 0.934 & 0.825 & 7.41 & 3.73 & 0.941 & 0.874 & 6.18 & 3.46 \\
        \midrule
\end{tabular}}
\label{tab:tab_results_uwb}
\end{table}

Fig.~\ref{fig:results_5G} visualizes the channel charts for the 5G data set. Graph (a) shows the reference radio environment with a color gradient, while the Graphs (b)--(f) show the results of the channel charts for different methods. We can clearly see that our method, cf.\ \ref{fig:5g_sg}, and the Isomap, cf.\ \ref{fig:5g_iso}, preserve the global structure of the environment very well. However, it overlaps slightly on the left-hand side and drifts away on the right-hand side on the top. As both methods rely on the nearest neighbors to create the geodesic distances, only vertical neighbors are available on the tail on the upper right-hand side, which leads to a overestimation of the Euclidean distance and therefore to a horizontal drift therein. The channel chart of the CAE~+~UMAP algorithm, cf.\ Fig.~\ref{fig:5g_umap}, does not reflect the global similarity very well, as its shape is curved. However, the local similarity is still reasonable as the color gradient is valid in most areas. The other algorithms have a much lower quality of the channel charts: The triplet approach, cf.\ Fig.~\ref{fig:5g_t}, and the constrained autoencoder, cf.\ Fig.~\ref{fig:5g_cond}, fail in recovering the geometry.

The generalization abilities are good for all of the algorithms, showing no significant differences of the performance using the test data set. This indicates that the algorithms are not overfitting to the training trajectories and can predict also well for unseen data in the same environment. 

\subsection{UWB}

\begin{figure*}[t]
    \centering
    \subfloat[Radio environment\label{fig:uwb_re}]{
        \includegraphics[trim={5mm 2mm 0 0}, width=.32\linewidth]{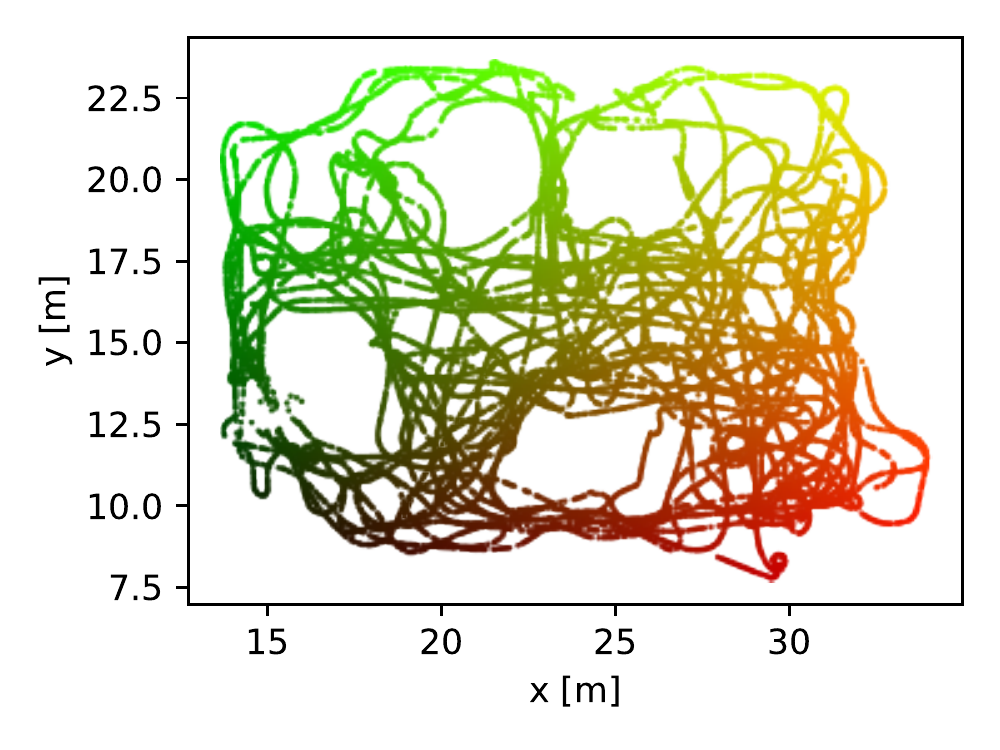}}
    \subfloat[Siamese Geo.\label{fig:uwb_sg}]{
        \includegraphics[trim={5mm 2mm 0 0},width=.32\linewidth]{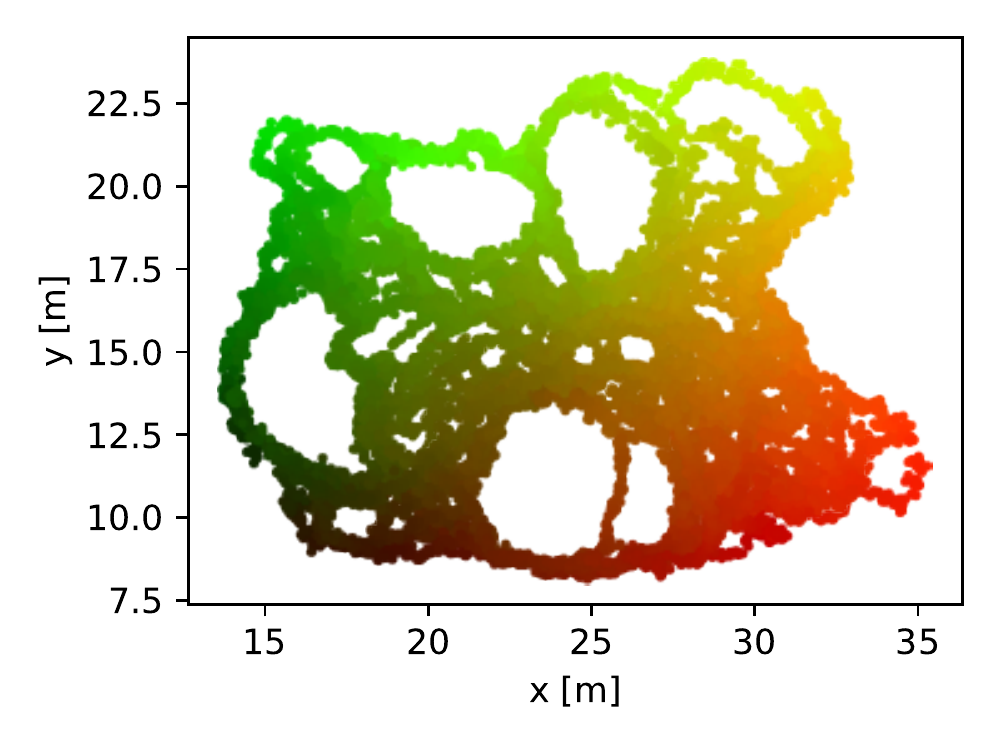}}
    \subfloat[CAE + UMAP \label{fig:uwb_umap}]{
        \includegraphics[trim={5mm 2mm 0 0},width=.32\linewidth]{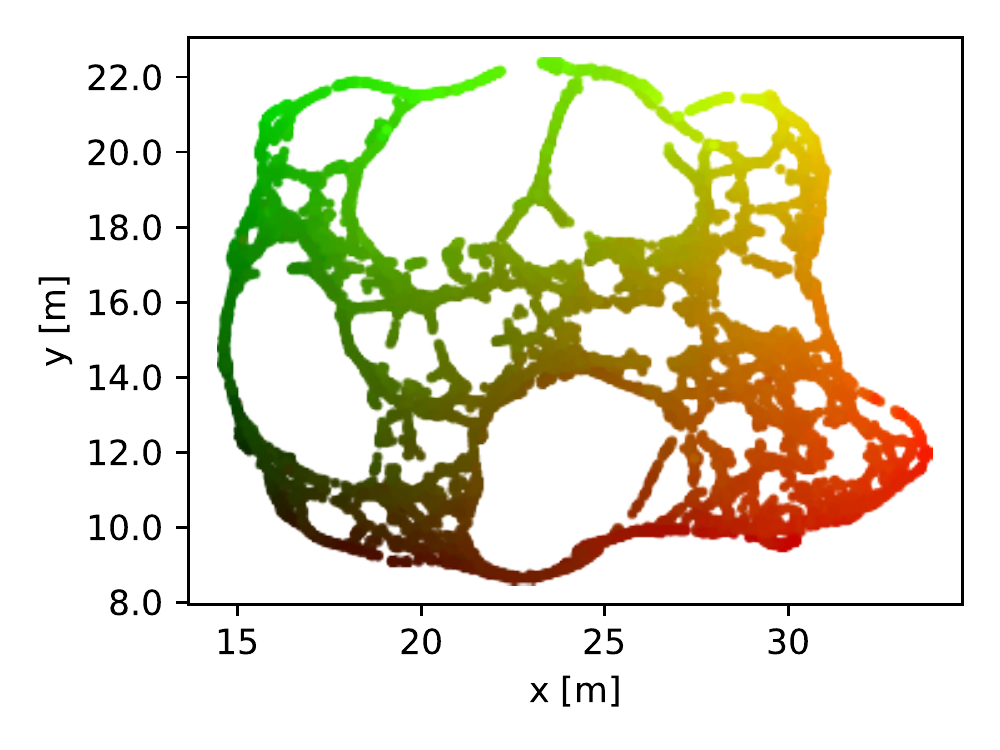}}
    \\
    \subfloat[Triplet\label{fig:uwb_t}]{
        \includegraphics[trim={5mm 2mm 0 0},width=.32\linewidth]{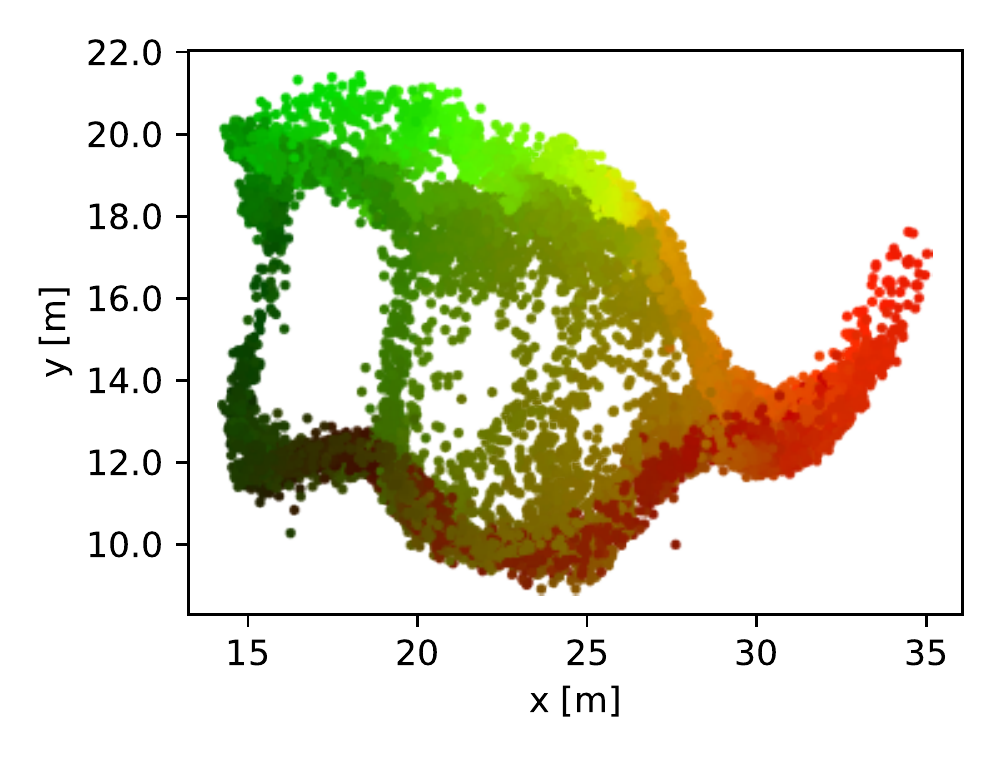}}
    \subfloat[Isomap\label{fig:uwb_iso}]{
        \includegraphics[trim={5mm 2mm 0 0},width=.32\linewidth]{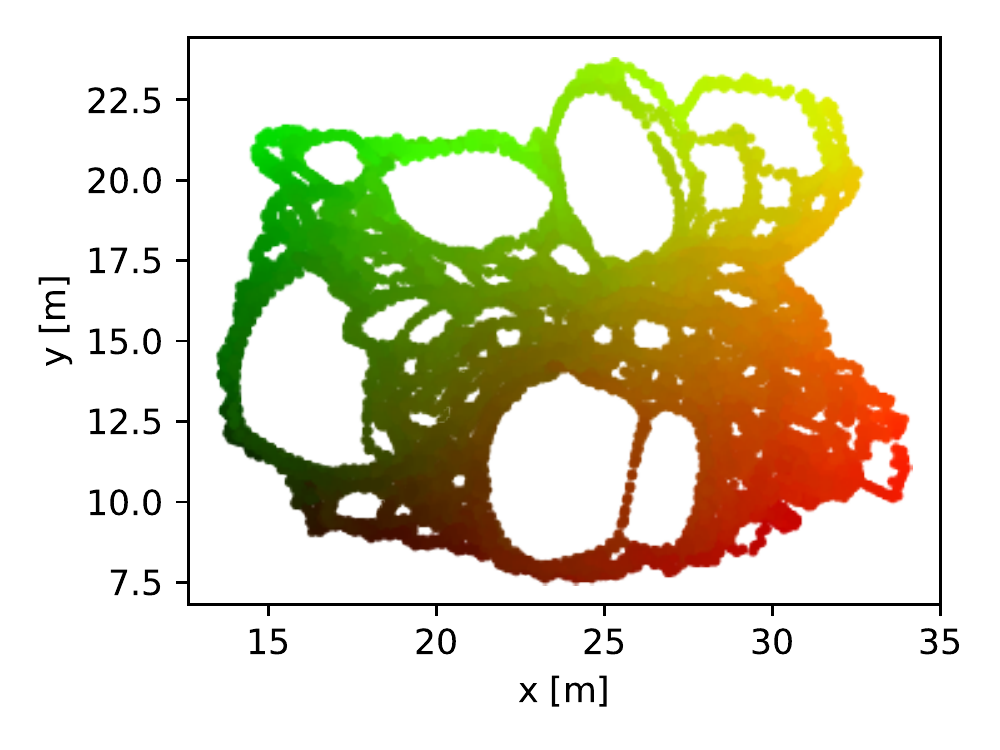}}
    \subfloat[Con.\ AE\label{fig:uwb_cond}]{
        \includegraphics[trim={5mm 2mm 0 0},width=.32\linewidth]{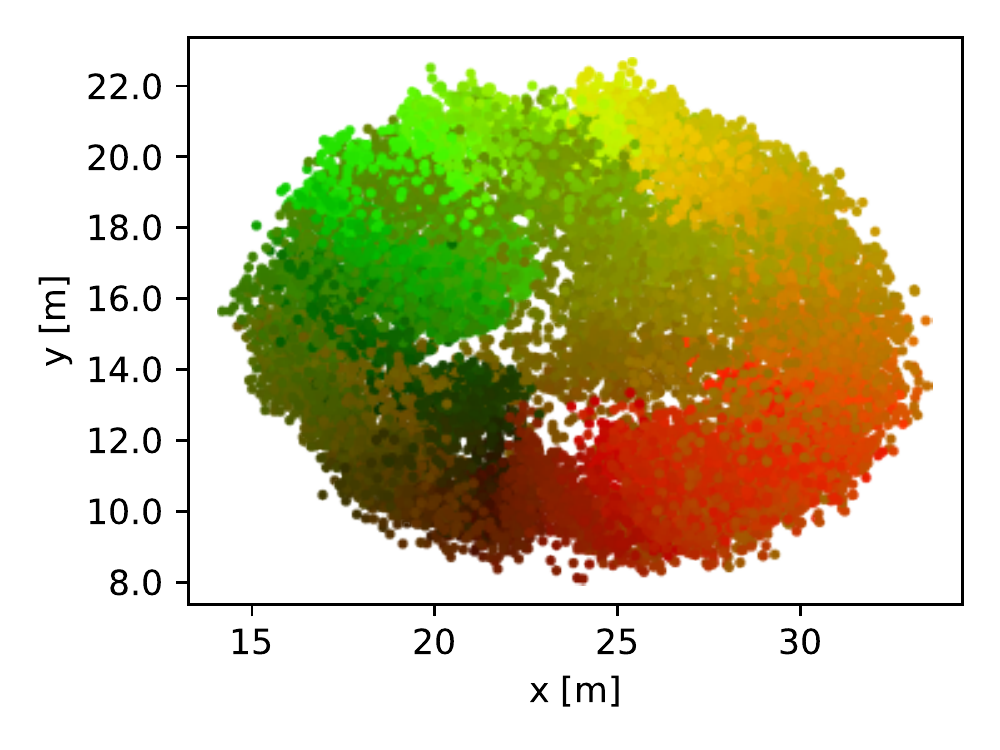}}
    \caption{Results of the CCs for the UWB data set. Graph (a) shows the recorded training data in the real-world environment, whereas the color gradient indicates the relation of the positions of the channel charts to the real positions. Graphs (b)--(f) show the channel charts after the affine transformation to the global coordinate frame.}
    \label{fig:results_uwb}
\end{figure*}

\noindent Table~\ref{tab:tab_results_uwb} lists the results of the CC methods on the UWB data set. The majority of the state-of-the-art CC methods show high accuracy for the local similarity with $\mathrm{TW}$ and $\mathrm{CT} > 0.95$ for our approach, Isomap, CAE~+~UMAP and the Triplet approach. Interestingly, CAE~+~UMAP achieves similarly high local similarity of $\mathrm{TW}$ and $\mathrm{CT} = 0.996$ and global similarity of $\mathrm{CE90} = 1.37\,\mathrm{m}$ and $\mathrm{MAE} = 0.72\,\mathrm{m}$ to ours and the Isomap. Also the triplet based algorithm achieves higher local accuracies of $\mathrm{TW} = 0.951$ and $\mathrm{CT} = 0.975$. We think the improvement comes from the slow and consistent movement of the robot platform with less dynamics in the movement pattern, which leads to a more valid time based triplet selection. However, recovering the global structure of the environment is still not possible with a weak global accuracy of ${\mathrm{CE90} = 3.98\,\mathrm{m}}$ and ${\mathrm{MAE} = 2.29\,\mathrm{m}}$. Like in the 5G evaluation, Con.~AE, Sammon mapping and PCA achieve the worst results in local and global similarity of ${\mathrm{TW} < 0.93}$, ${\mathrm{CT} > 0.97}$, ${\mathrm{CE90} > 4.70\,\mathrm{m}}$ and ${\mathrm{MAE} > 2.50\,\mathrm{m}}$. 

Fig.~\ref{fig:results_uwb} visualizes the channel charts for the UWB data set. Graph (a) shows the reference radio environment with a color gradient, while the Graphs (b)--(f) show the results of the channel charts for different methods. We can clearly see that our method and Isomap model the global structure of the environment well. There is no overlap of the gradients and the majority of the empty areas in the original environment are also present. However, due to the empty areas in the environment, the geodesic distances are overestimated as no direct Euclidean connection via the neighborhood is possible for all data points, which leads to several over-estimations of the geodesic distance to the Euclidean distances. The effect can be seen in the area at $x=15\,\mathrm{m}$ and $y=15\,\mathrm{m}$, where the channel-chart coordinates are pushed towards the left-hand side. The channel chart generated by CAE~+~UMAP is similarly accurate as the one generated by our approach and Isomap. The chart is smoother but maps near datapoints on similar trajectories, with some discontinuities. In contrast, our geodesic Siamese network shows no discontinuities of the trajectories. While CAE + UMAP can achieve a globally consistent map in the UWB scenario UMAP does not ensure this behaviour as shown in the 5G evaluation, cf.\ \ref{fig:5g_umap}. We therefore think that CAE~+~UMAP is not ideal for CC-assisted localization, which needs a continuous globally consistent chart. The other channel charts, i.e., triplet and Con.~AE, cannot recover the global geometry well, while the triplet algorithm provides more structure of the environment. This may be caused by the restrictive assumptions of the triplet generation and the limitations of the distance metric for far points for Con.~AE.

Also here, the generalization abilities are good for all of the algorithms, showing no significant differences of the performance using the test data set. This indicates that the algorithms are not overfitting to the training trajectories and can predict also well for unseen data in the same environment. 

\section{Discussion \& Future Work}
\label{section:discussion}

\noindent The experiments have shown that neighborhood-graph-based algorithms, e.g., CAE~+~UMAP, Isomap, and the proposed geodesic Siamese network, are superior in the generation of channel charts compared to the other algorithms. We think that this is due to the fact that they only consider local distances, which is advantageous as the CSI distance is proportional to the Euclidean distance in the proximity of two CSI measurements. This leads to the optimization of the CC considering only the valid local connections between CSIs. However, this also means that, ideally, a uniform distribution of high data density should be available to create a valid global neighborhood graph. As the idea of CC is that every user in the environment can contribute to the underlying data set, the amount of collected data will grow rapidly, which leads to a high data density. However, this does not necessarily imply that the density of data has a uniform distribution in the radio environment. While this is not a problem, as long as the data density is high, methods that rely on proximity distances, such as nearest-neighbor approaches, have to calculate a distance matrix, which has a complexity of $\mathcal{O}(n^2)$. 

While there are good approximations for neighborhood graphs \cite{wang2021comprehensive} the complexity is still non-linear, which restricts the most promising algorithms to small data sets. The selection of the most necessary information of the recorded data is therefore important, but a non-trivial problem, as a random selection of samples might lead to redundant CSIs at high density regions of the environment and therefore also to a under representation of low-density regions. As the collection of data is unsupervised, a spatial selection of data is impossible, which makes the sample selection challenging.

Another challenge is to transform the local channel chart into the global coordinate frame, which is necessary for localization. There are several semi-supervised approaches solving this problem, by adding a constraint to the non-linear CC optimization to transform the local chart into the real-world frame \cite{lei2019siamese, ferrand2021triplet, deng2021network, zhang2021semi}. This solves the scaling problem of the approximate distance metrics, the arbitrary rotation of the local channel chart and its non-linear deformation such as overlapping areas. However, if a channel chart does not have any deformations, a simple linear affine transformation is optimal to match its coordinate frame and therefore only needs three labeled channel measurements in the real world. In contrast, by adding a constraint into the CC objective, typically a larger amount of data is needed as the algorithms try to learn a non-linear, highly flexible transformation and therefore tend to overfit to the data \cite{lawrence2000overfitting}. We therefore recommend a two-step optimization approach for localization, by i) learning the geometry of the radio environment in an unsupervised way and ii) to transform the learned model into the global coordinate frame by a simple linear transformation. This leads to the lowest amount of needed labeled data if the channel chart does reflect its geometry well enough.

A further problem is that the consistency of the CC mainly depends on the data collected in a certain scenario. Environmental changes may alter the environment-specific CSI, which leads to errors in the prediction. This is a common problem in fingerprinting models \cite{stahlke20225g} and makes an identification of environmental changes and an update necessary over time. This might also be needed for channel charts rendering a CC life-cycle management essential for a robust usage for downstream tasks.

\section{Conclusion}
\label{section:conclusion}

\noindent In this work, we propose a channel charting method for synchronized multi-base-station SISO radio setups with high bandwidth, a 5G and a UWB setup. We derived a novel CSI distance metric, which allows to model globally consistent channel charts enabling high precision fingerprinting without the need for ground-truth label information. Our proposed method, based on a Siamese network can achieve localization accuracies of $0.69\,\mathrm{m}$ for the UWB and $1.4\,\mathrm{m}$ for the 5G setup in challenging NLoS dominated indoor environments. An extensive study has shown that our proposed methods outperforms the state-of-the-art algorithms, while neighborhood-graph-based algorithms like our method can achieve the best results. This indicates that CC has a very high potential for unsupervised fingerprinting-based localization methods to lower the effort for data recording, as only very few ground-truth data are needed for the coordinate-system transformation of the chart from the local to the real-world frame.

\appendices
\section{Model for TDoA-Based Measurements}
\label{appendix:tdoa}

\noindent The CIR distance for TDoA-based localization system is derived in the following. Without loss of generality, we assume $B_\mathsf{ref}$ as the time-reference base station for TDoA-based measurements in the following. Moreover, $B_\mathsf{ref}$ is always chosen such that it provides the shortest ToA. The difference between two TDoA measurements ($T^{}_{\mathsf{bias}, \ell}$ and $T^{}_{\mathsf{bias}, \ell^\prime}$ are some measurement bias) is given by
\begingroup
\allowdisplaybreaks
\begin{align}
    \Delta T_{n,i \cs j}^{(k)} &\defeq \big| T_{0,i}^{B_k} + T^{}_{\mathsf{bias}, \ell} - ( T_{0,i}^{B_\mathsf{ref}} + T^{}_{\mathsf{bias}, \ell}) \nonumber \\ 
    &\qquad -
    ( T_{0,j}^{B_k} + T^{}_{\mathsf{bias}, \ell^\prime} - ( T_{0,j}^{B_\mathsf{ref}} + T^{}_{\mathsf{bias}, \ell^\prime}) ) \big| \nonumber \\
    &= \big| T_{0,i}^{B_k} - T_{0,i}^{B_\mathsf{ref}} - (T_{0,j}^{B_k} - T_{0,j}^{B_\mathsf{ref}}) \big| \nonumber \\ 
    &=  \big| T_{0,i}^{B_k} -  T_{0,j}^{B_k} + T_{0,j}^{B_\mathsf{ref}} - T_{0,i}^{B_\mathsf{ref}} \big| \nonumber \\ 
    &= \big| \Delta t^{B_k}_{n,i \cs j} + \Delta t^{B_\mathsf{ref}}_{n,j \cs i} \big|  \;, \nonumber \label{eq:TDOA} \\
  \intertext{which can be expressed as}
&= \left| \pm \sqrt{ \left(d_\mathsf{euc}^{(k)} \right)^2 / \mathrm{c}^2 - \left(\epsilon_{n,i \cs j}^{(k)}\right)^2} \right. \nonumber \\ &\qquad \qquad \left. \pm \sqrt{ \left(d_\mathsf{euc}^{(\mathsf{ref})} \right)^2 / \mathrm{c}^2 - \left(\epsilon_{n,i \cs j}^{(\mathsf{ref})}\right)^2} \right| \;.
\end{align}
\endgroup
When $\epsilon_{n,i \cs j}^{(k)}$ is maximum, then $\Delta T \in [0, d_\mathsf{euc}/\mathrm{c}]$ which implies $\epsilon_{n,i \cs j}^{(\mathsf{ref})} \in [0, d_\mathsf{euc}/\mathrm{c}]$. Similarly, when $\epsilon_{n,i \cs j}^{(\mathsf{ref})}$ is maximum,  $\epsilon_{n,i \cs j}^{(k)} \in [0, d_\mathsf{euc}/\mathrm{c}]$. This means that
\begin{align}
     2 \, d_\mathsf{euc} > \mathrm{c}\, \big(\epsilon_{n,i \cs j}^{(k)} + \epsilon_{n,i \cs j}^{(\mathsf{ref})} \big) \;, \label{eq:app_lower_bound}
\end{align}
since equality in \eqref{eq:app_lower_bound} can only hold when both base stations are at the exact same coordinates.

Following the same derivation as in Section~\ref{chap:cir_dist}, we now re-define the distance metric \eqref{eq:distance_metric}  using TDoAs, i.e., $\Delta T_{n,i \cs j}^{(k)}$:
\begin{align}
      d_\mathsf{s}^{}(\ve{\Tilde{h}}_i^{}, \ve{\Tilde{h}}_j^{}) \defeq \sum_{k=0}^{N_\mathsf{b}-1} \sum_{n=0}^{N_\mathsf{p}-1} \Delta T_{n,i \cs j}^{(k)} \;.
\end{align}
Again, we have a monotonically increasing mean and standard deviation, and a lower bound of the error.

\section{Linearity of CIR distance}
\label{appendix:cir_linear}

\noindent If we could extract all multipath delay information from the received CSIs we could apply \eqref{eq:distance_metric}, which would lead to a linear correlation of the CIR distance to the Euclidean distance. In this work, we use the approximation defined in \eqref{eq:dist_approx}, which is non-linear if we use a typical pulse shape like a sinc function. However, we can show that the approximated CIR distance is linear for small time-differences $\Delta t^{(k)}_{n,i \cs j}$. Without loss of generality, we assume to have only one base station $N_\mathsf{b} = 1$ and a single path arriving at the receiver $N_\mathsf{p} = 1$. By using \eqref{eq:dist_approx} and the inverse triangle inequality we have
\begin{align} \label{eq:dist}
    d_\mathsf{s}^{}(\ve{\Tilde{h}}_i^{}, \ve{\Tilde{h}}_j^{}) &\leq \sum_{t=0}^{T-1} \big| \Tilde{h}^{(1)}_i(t) - \Tilde{h}^{(1)}_j(t) \big| \;.
    \intertext{If we assume to have only a single path arriving, e.g., the LoS component, we can substitute $\Tilde{h}$ by the sinc function}
    d_\mathsf{s}^{}(\ve{\Tilde{h}}_i^{}, \ve{\Tilde{h}}_j^{}) &\leq \sum_{t=0}^{T-1} \left| \sinc( t )  - \sinc( t - \Delta t^{(1)}_{0,i \cs j}) \right| \;.
    \intertext{Here, we omit the bandwidth of the signal and assume to have a equal amplitude, which is valid for small $\Delta t^{(1)}_{0,i \cs j}$. By using the Taylor series expansion we get}
    d_\mathsf{s}^{}(\ve{\Tilde{h}}_i^{}, \ve{\Tilde{h}}_j^{}) &\leq \sum_{t=0}^{T-1} \left| \sum_{n=0}^\infty \frac{(-1)^n t^{2n}}{(2n+1)!} \right. \nonumber \\
    &\qquad \left. - \sum_{n=0}^\infty \frac{(-1)^n (t-\Delta t^{(1)}_{0,i \cs j})^{2n}}{(2n+1)!} \right| \;.
    \intertext{If we have $|\Delta t^{(1)}_{0,i \cs j}| \ll |t|$ we can use the binomial approximation defined in Lemma \ref{lem:bio_approx} to get}
   d_\mathsf{s}^{}(\ve{\Tilde{h}}_i^{}, \ve{\Tilde{h}}_j^{}) &\leq \sum_{t=0}^{T-1} \left| \sum_{n=0}^\infty \frac{(-1)^n t^{2n}}{(2n+1)!} \right. \nonumber \\ 
    &\qquad \left. - \sum_{n=0}^\infty  \frac{(-1)^n (t^{2n} - 2n t^{2n-1} \Delta t^{(1)}_{0,i \cs j}) }{(2n+1)!} \right| \\
    &= \left|\Delta t^{(1)}_{0,i \cs j}\right| \underbrace{\sum_{t=0}^{T-1} \left|  \sum_{n=0}^\infty \frac{(-1)^n 2n t^{2n-1} }{(2n+1)!} \right|}_{=\text{constant}} \;,
\end{align}
which is linear for small $\Delta t^{(1)}_{0,i \cs j}$ with a constant slope. This means that also the sum of $\Delta t^{(k)}_{n,i \cs j}$ for the base stations $k$ and the received paths $n$ is linear, as long as $\Delta t^{(k)}_{n,i \cs j}$ is small. As $\Delta t^{(k)}_{n,i \cs j}$ is bounded by the displacements of the CSI measurements $\ve{\Tilde{h}}_i^{}$ and $\ve{\Tilde{h}}_j^{}$, we have a linear CIR distance for small displacements.

\begin{lemma} \label{lem:bio_approx}
For $|\Delta t| \ll |t|$ we can use the following approximation of the binomial function
    \begin{align}
        f(\Delta t) &= (t - \Delta t)^{2n} \\
        f'(\Delta t) &= -2n(t-\Delta t)^{{2n}-1} \\
        f'(0) &= -2n t^{{2n}-1} \\
        f(\Delta t) &\approx f(0) + f'(0) (\Delta t-0) \\
            &= t^{2n} - {2n} t^{{2n}-1} \Delta t \;.
    \end{align}
    The error of the approximation is defined as 
    \begin{align}
        e(\Delta t) = (t-\Delta t)^{2n} - (t^{2n} - {2n} t^{{2n}-1} \Delta t) \;,
    \end{align}
    where for small values of $\Delta t$, the error approaches $0$, i.e.,
    \begin{align}
        \lim_{\Delta t \to 0} e(\Delta t) = 0 \;.
    \end{align}
\end{lemma}

\bibliographystyle{IEEEtranM} 
\bibliography{bibliography}%

\begin{thebibliography}{10}
\providecommand{\url}[1]{#1}
\csname url@samestyle\endcsname
\providecommand{\newblock}{\relax}
\providecommand{\bibinfo}[2]{#2}
\providecommand{\BIBentrySTDinterwordspacing}{\spaceskip=0pt\relax}
\providecommand{\BIBentryALTinterwordstretchfactor}{4}
\providecommand{\BIBentryALTinterwordspacing}{\spaceskip=\fontdimen2\font plus
\BIBentryALTinterwordstretchfactor\fontdimen3\font minus
  \fontdimen4\font\relax}
\providecommand{\BIBforeignlanguage}[2]{{%
\expandafter\ifx\csname l@#1\endcsname\relax
\typeout{** WARNING: IEEEtran.bst: No hyphenation pattern has been}%
\typeout{** loaded for the language `#1'. Using the pattern for}%
\typeout{** the default language instead.}%
\else
\language=\csname l@#1\endcsname
\fi
#2}}
\providecommand{\BIBdecl}{\relax}
\BIBdecl

\bibitem{laoudias2018survey}
C.~Laoudias, A.~Moreira, S.~Kim, S.~Lee, L.~Wirola, and C.~Fischione, ``{A
  Survey of Enabling Technologies for Network Localization, Tracking, and
  Navigation},'' \emph{IEEE Communications Surveys \& Tutorials}, vol.~20,
  no.~4, pp. 3607--3644, 2018.

\bibitem{wu2018image}
Y.~Wu, F.~Tang, and H.~Li, ``Image-based camera localization: an overview,''
  \emph{Visual Computing for Industry, Biomedicine, and Art}, vol.~1, no.~8,
  2018.

\bibitem{elhousni2020survey}
M.~Elhousni and X.~Huang, ``{A Survey on 3D LiDAR Localization for Autonomous
  Vehicles},'' in \emph{IEEE Intelligent Vehicles Symposium (IV)}.\hskip 1em
  plus 0.5em minus 0.4em\relax Las Vegas, NV, USA: IEEE, 2020, pp. 1879--1884.

\bibitem{rahman2020recent}
A.~M. Rahman, T.~Li, and Y.~Wang, ``{Recent Advances in Indoor Localization via
  Visible Lights: A Survey},'' \emph{Sensors}, vol.~20, no.~5, p. 1382, 2020.

\bibitem{saeed2019state}
N.~Saeed, H.~Nam, T.~Y. Al-Naffouri, and M.-S. Alouini, ``{A State-of-the-Art
  Survey on Multidimensional Scaling-Based Localization Techniques},''
  \emph{IEEE Communications Surveys \& Tutorials}, vol.~21, no.~4, pp.
  3565--3583, 2019.

\bibitem{pang2020aoa}
F.~Pang, K.~Do{\u{g}}an{\c{c}}ay, N.~H. Nguyen, and Q.~Zhang, ``{AOA
  Pseudolinear Target Motion Analysis in the Presence of Sensor Location
  Errors},'' \emph{IEEE Transactions on Signal Processing}, vol.~68, pp.
  3385--3399, 2020.

\bibitem{gifford2020impact}
W.~Gifford, D.~Dardari, and M.~Win, ``{The Impact of Multipath Information on
  Time-of-Arrival Estimation},'' \emph{IEEE Transactions on Signal Processing},
  vol.~70, pp. 31--46, 2020.

\bibitem{christopher2021characterizing}
C.~E. O’Lone, H.~S. Dhillon, and R.~M. Buehrer, ``{Characterizing the
  First-Arriving Multipath Component in 5G Millimeter Wave Networks: TOA, AOA,
  and Non-Line-of-Sight Bias},'' \emph{IEEE Transactions on Wireless
  Communications}, vol.~21, no.~3, pp. 1602--1620, 2021.

\bibitem{stahlke2020nlos}
M.~Stahlke, S.~Kram, C.~Mutschler, and T.~Mahr, ``{NLOS Detection using UWB
  Channel Impulse Responses and Convolutional Neural Networks},'' in
  \emph{International Conference on Localization and GNSS (ICL-GNSS)}.\hskip
  1em plus 0.5em minus 0.4em\relax Tampere, Finland: IEEE, 2020.

\bibitem{stahlke2021estimating}
M.~Stahlke, S.~Kram, F.~Ott, T.~Feigl, and C.~Mutschler, ``{Estimating TOA
  Reliability With Variational Autoencoders},'' \emph{IEEE Sensors Journal},
  vol.~22, no.~6, pp. 5133--5140, 2021.

\bibitem{niitsoo2019deep}
A.~Niitsoo, T.~Edelh{\"a}u{\ss}er, E.~Eberlein, N.~Hadaschik, and C.~Mutschler,
  ``{A Deep Learning Approach to Position Estimation from Channel Impulse
  Responses},'' \emph{Sensors}, vol.~19, no.~5, p. 1064, 2019.

\bibitem{stahlke20225g}
M.~Stahlke, T.~Feigl, M.~H.~C. García, R.~A. Stirling-Gallacher, J.~Seitz, and
  C.~Mutschler, ``{Transfer Learning to adapt 5G AI-based Fingerprint
  Localization across Environments},'' in \emph{IEEE 95th Vehicular Technology
  Conference (VTC2022-Spring)}.\hskip 1em plus 0.5em minus 0.4em\relax
  Helsinki, Finland: IEEE, 2022.

\bibitem{liu2017toward}
K.~Liu, H.~Zhang, J.~K.-Y. Ng, Y.~Xia, L.~Feng, V.~C. Lee, and S.~H. Son,
  ``{Toward Low-Overhead Fingerprint-Based Indoor Localization via Transfer
  Learning: Design, Implementation, and Evaluation},'' \emph{IEEE Transactions
  on Industrial Informatics}, vol.~14, no.~3, pp. 898--908, 2017.

\bibitem{de2020csi}
S.~De~Bast, A.~P. Guevara, and S.~Pollin, ``{CSI-based Positioning in Massive
  MIMO systems using Convolutional Neural Networks},'' in \emph{IEEE 91st
  Vehicular Technology Conference (VTC2020-Spring)}.\hskip 1em plus 0.5em minus
  0.4em\relax Antwerp, Belgium: IEEE, 2020.

\bibitem{widmaier2019towards}
M.~Widmaier, M.~Arnold, S.~Dorner, S.~Cammerer, and S.~ten Brink, ``{Towards
  Practical Indoor Positioning Based on Massive MIMO Systems},'' in \emph{IEEE
  90th Vehicular Technology Conference (VTC2019-Fall)}.\hskip 1em plus 0.5em
  minus 0.4em\relax Honolulu, HI, USA: IEEE, 2019.

\bibitem{studer2018channel}
C.~Studer, S.~Medjkouh, E.~Gonulta{\c{s}}, T.~Goldstein, and O.~Tirkkonen,
  ``{Channel Charting: Locating Users Within the Radio Environment Using
  Channel State Information},'' \emph{IEEE Access}, vol.~6, pp.
  47\,682--47\,698, 2018.

\bibitem{lei2019siamese}
E.~Lei, O.~Casta{\~n}eda, O.~Tirkkonen, T.~Goldstein, and C.~Studer, ``{Siamese
  Neural Networks for Wireless Positioning and Channel Charting},'' in
  \emph{57th Annual Allerton Conference on Communication, Control, and
  Computing (Allerton)}.\hskip 1em plus 0.5em minus 0.4em\relax Monticello, IL,
  USA: IEEE, 2019, pp. 200--207.

\bibitem{ferrand2021triplet}
P.~Ferrand, A.~Decurninge, L.~G. Ordo{\~n}ez, and M.~Guillaud, ``{Triplet-Based
  Wireless Channel Charting: Architecture and Experiments},'' \emph{IEEE
  Journal on Selected Areas in Communications}, vol.~39, no.~8, pp. 2361--2373,
  2021.

\bibitem{deng2021network}
J.~Deng, O.~Tirkkonen, J.~Zhang, X.~Jiao, and C.~Studer, ``{Network-side
  Localization via Semi-Supervised Multi-point Channel Charting},'' in
  \emph{International Wireless Communications and Mobile Computing
  (IWCMC)}.\hskip 1em plus 0.5em minus 0.4em\relax Harbin City, China: IEEE,
  2021, pp. 1654--1660.

\bibitem{zhang2021semi}
Q.~Zhang and W.~Saad, ``{Semi-Supervised Learning for Channel Charting-Aided
  IoT Localization in Millimeter Wave Networks},'' in \emph{IEEE Global
  Communications Conference (GLOBECOM)}.\hskip 1em plus 0.5em minus 0.4em\relax
  Madrid, Spain: IEEE, 2021.

\bibitem{ribeiro2020channel}
L.~Ribeiro, M.~Leinonen, H.~Djelouat, and M.~Juntti, ``{Channel Charting for
  Pilot Reuse in mMTC with Spatially Correlated MIMO Channels},'' in \emph{IEEE
  Globecom Workshops}.\hskip 1em plus 0.5em minus 0.4em\relax Taipei, Taiwan:
  IEEE, 2020.

\bibitem{al2021adaptive}
H.~Al-Tous, O.~Tirkkonen, and J.~Liang, ``{Adaptive Sector Splitting based on
  Channel Charting in Massive MIMO Cellular Systems},'' in \emph{IEEE 93rd
  Vehicular Technology Conference (VTC2021-Spring)}.\hskip 1em plus 0.5em minus
  0.4em\relax Helsinki, Finland: IEEE, 2021.

\bibitem{al2020multipoint}
H.~Al-Tous, T.~Ponnada, C.~Studer, and O.~Tirkkonen, ``Multipoint channel
  charting-based radio resource management for v2v communications,''
  \emph{EURASIP Journal on Wireless Communications and Networking}, no. 132,
  2020.

\bibitem{ponnada2021location}
T.~Ponnada, H.~Al-Tous, and O.~Tirkkonen, ``{Location-Free Beam Prediction in
  mmWave Systems},'' in \emph{IEEE 93rd Vehicular Technology Conference
  (VTC2021-Spring)}.\hskip 1em plus 0.5em minus 0.4em\relax Helsinki, Finland:
  IEEE, 2021, pp. 1--6.

\bibitem{ponnada2021best}
T.~Ponnada, P.~Kazemi, H.~Al-Tous, Y.-C. Liang, and O.~Tirkkonen, ``{Best Beam
  Prediction in Non-Standalone mm Wave Systems},'' in \emph{Joint European
  Conference on Networks and Communications \& 6G Summit (EuCNC/6G
  Summit)}.\hskip 1em plus 0.5em minus 0.4em\relax Porto, Portugal: IEEE, 2021,
  pp. 532--537.

\bibitem{kazemi2021channel}
P.~Kazemi, T.~Ponnada, H.~Al-Tous, Y.-C. Liang, and O.~Tirkkonen, ``{Channel
  Charting Based Beam SNR Prediction},'' in \emph{Joint European Conference on
  Networks and Communications \& 6G Summit (EuCNC/6G Summit)}.\hskip 1em plus
  0.5em minus 0.4em\relax Porto, Portugal: IEEE, 2021, pp. 72--77.

\bibitem{deng2018multipoint}
J.~Deng, S.~Medjkouh, N.~Malm, O.~Tirkkonen, and C.~Studer, ``{Multipoint
  Channel Charting for Wireless Networks},'' in \emph{52nd Asilomar Conference
  on Signals, Systems, and Computers}.\hskip 1em plus 0.5em minus 0.4em\relax
  Pacific Grove, CA, USA: IEEE, 2018, pp. 286--290.

\bibitem{le2021efficient}
L.~Le~Magoarou, ``{Efficient Channel Charting via Phase-Insensitive Distance
  Computation},'' \emph{IEEE Wireless Communications Letters}, vol.~10, no.~12,
  pp. 2634--2638, 2021.

\bibitem{agostini2020channel}
P.~Agostini, Z.~Utkovski, and S.~Sta{\'n}czak, ``{Channel Charting: an
  Euclidean Distance Matrix Completion Perspective},'' in \emph{IEEE
  International Conference on Acoustics, Speech and Signal Processing
  (ICASSP)}.\hskip 1em plus 0.5em minus 0.4em\relax Barcelona, Spain: IEEE,
  2020, pp. 5010--5014.

\bibitem{schmidt1986multiple}
R.~Schmidt, ``{Multiple Emitter Location and Signal Parameter Estimation},''
  \emph{IEEE Transactions on Antennas and Propagation}, vol.~34, no.~3, pp.
  276--280, 1986.

\bibitem{spielman1986music}
D.~Spielman, A.~Paulraj, and T.~Kailath, ``{Performance Analysis of the MUSIC
  Algorithm},'' in \emph{IEEE International Conference on Acoustics, Speech,
  and Signal Processing (ICASSP)}, vol.~11.\hskip 1em plus 0.5em minus
  0.4em\relax Tokyo, Japan: IEEE, 1986, pp. 1909--1912.

\bibitem{ponnada2019out}
T.~Ponnada, H.~Al-Tous, O.~Tirkkonen, and C.~Studer, ``{An Out-of-Sample
  Extension for Wireless Multipoint Channel Charting},'' in \emph{International
  Conference on Cognitive Radio Oriented Wireless Networks}.\hskip 1em plus
  0.5em minus 0.4em\relax Poznan, Poland: Springer, 2019, pp. 208--217.

\bibitem{geng2020multipoint}
C.~Geng, H.~Huang, and J.~Langerman, ``{Multipoint Channel Charting With
  Multiple-Input Multiple-Output Convolutional Autoencoder},'' in
  \emph{IEEE/ION Position, Location and Navigation Symposium (PLANS)}.\hskip
  1em plus 0.5em minus 0.4em\relax Portland, OR, USA: IEEE, 2020, pp.
  1022--1028.

\bibitem{huang2019improving}
P.~Huang, O.~Casta{\~n}eda, E.~G{\"o}n{\"u}lta{\c{s}}, S.~Medjkouh,
  O.~Tirkkonen, T.~Goldstein, and C.~Studer, ``{Improving Channel Charting with
  Representation-Constrained Autoencoders},'' in \emph{IEEE 20th International
  Workshop on Signal Processing Advances in Wireless Communications
  (SPAWC)}.\hskip 1em plus 0.5em minus 0.4em\relax Cannes, France: IEEE, 2019.

\bibitem{ferrand2020triplet}
P.~Ferrand, A.~Decurninge, L.~G. Ordo{\~n}ez, and M.~Guillaud, ``{Triplet-Based
  Wireless Channel Charting},'' in \emph{IEEE Global Communications Conference
  (GLOBECOM)}.\hskip 1em plus 0.5em minus 0.4em\relax Taipei, Taiwan: IEEE,
  2020.

\bibitem{rappaport2021improving}
B.~Rappaport, E.~G{\"o}n{\"u}lta{\c{s}}, J.~Hoydis, M.~Arnold, P.~K. Srinath,
  and C.~Studer, ``{Improving Channel Charting using a Split Triplet Loss and
  an Inertial Regularizer},'' in \emph{17th International Symposium on Wireless
  Communication Systems (ISWCS)}.\hskip 1em plus 0.5em minus 0.4em\relax
  Berlin, Germany: IEEE, 2021.

\bibitem{euchner2022improving}
F.~Euchner, P.~Stephan, M.~Gauger, S.~D{\"o}rner, and S.~Ten~Brink,
  ``{Improving Triplet-Based Channel Charting on Distributed Massive MIMO
  Measurements},'' in \emph{IEEE 23rd International Workshop on Signal
  Processing Advances in Wireless Communication (SPAWC)}.\hskip 1em plus 0.5em
  minus 0.4em\relax Oulu, Finland: IEEE, 2022.

\bibitem{agostini2022not}
P.~Agostini, Z.~Utkovski, S.~Sta{\'n}czak, A.~A. Memon, B.~Zafar, and
  M.~Haardt, ``{Not-Too-Deep Channel Charting (N2D-CC)},'' in \emph{IEEE
  Wireless Communications and Networking Conference (WCNC)}.\hskip 1em plus
  0.5em minus 0.4em\relax Austin, TX, USA: IEEE, 2022, pp. 2160--2165.

\bibitem{pihlajasalo2020absolute}
J.~Pihlajasalo, M.~Koivisto, J.~Talvitie, S.~Ali-L{\"o}ytty, and M.~Valkama,
  ``{Absolute Positioning with Unsupervised Multipoint Channel Charting for 5G
  Networks},'' in \emph{IEEE 92nd Vehicular Technology Conference
  (VTC2020-Fall)}.\hskip 1em plus 0.5em minus 0.4em\relax Victoria, BC, Canada:
  IEEE, 2020.

\bibitem{kram2022delay}
S.~Kram, C.~Kraus, M.~Stahlke, T.~Feigl, J.~Thielecke, and C.~Mutschler,
  ``{Delay Estimation in Dense Multipath Environments using Time Series
  Segmentation},'' in \emph{IEEE Wireless Communications and Networking
  Conference (WCNC)}.\hskip 1em plus 0.5em minus 0.4em\relax Austin, TX, USA:
  IEEE, 2022, pp. 1671--1676.

\bibitem{tenenbaum2000global}
J.~B. Tenenbaum, V.~d. Silva, and J.~C. Langford, ``{A Global Geometric
  Framework for Nonlinear Dimensionality Reduction},'' \emph{Science}, vol.
  290, no. 5500, pp. 2319--2323, 2000.

\bibitem{dijkstra1959note}
E.~W. Dijkstra, ``{A Note on Two Problems in Connexion with Graphs},''
  \emph{Numerische mathematik}, vol.~1, no.~1, pp. 269--271, 1959.

\bibitem{Kirk1970}
D.~E. Kirk, \emph{Optimal Control: An Introduction}.\hskip 1em plus 0.5em minus
  0.4em\relax Prentice-Hall, 1970.

\bibitem{de2004sparse}
V.~De~Silva and J.~B. Tenenbaum, ``Sparse multidimensional scaling using
  landmark points,'' Technical report, Stanford University, Tech. Rep., 2004.

\bibitem{pai2019dimal}
G.~Pai, R.~Talmon, A.~Bronstein, and R.~Kimmel, ``{DIMAL: Deep Isometric
  Manifold Learning Using Sparse Geodesic Sampling},'' in \emph{IEEE Winter
  Conference on Applications of Computer Vision (WACV)}.\hskip 1em plus 0.5em
  minus 0.4em\relax Waikoloa, HI, USA: IEEE, 2019, pp. 819--828.

\bibitem{chicco2021siamese}
D.~Chicco, ``{Siamese Neural Networks: An Overview},'' \emph{Artificial Neural
  Networks}, pp. 73--94, 2021.

\bibitem{ismail2019deep}
H.~Ismail~Fawaz, G.~Forestier, J.~Weber, L.~Idoumghar, and P.-A. Muller, ``Deep
  learning for time series classification: a review,'' \emph{Data mining and
  knowledge discovery}, vol.~33, no.~4, pp. 917--963, 2019.

\bibitem{wang2021comprehensive}
M.~Wang, X.~Xu, Q.~Yue, and Y.~Wang, ``{A Comprehensive Survey and Experimental
  Comparison of Graph-Based Approximate Nearest Neighbor Search},'' \emph{Proc.
  VLDB Endow.}, vol.~14, no.~11, p. 1964–1978, jul 2021.

\bibitem{lawrence2000overfitting}
S.~Lawrence and C.~L. Giles, ``{Overfitting and Neural Networks: Conjugate
  Gradient and Backpropagation},'' in \emph{International Joint Conference on
  Neural Networks (IJCNN)}, vol.~1.\hskip 1em plus 0.5em minus 0.4em\relax
  Como, Italy: IEEE, 2000, pp. 114--119.

\end{thebibliography}

\begin{IEEEbiography}[{\includegraphics[width=1in,height=1.25in,clip,keepaspectratio]{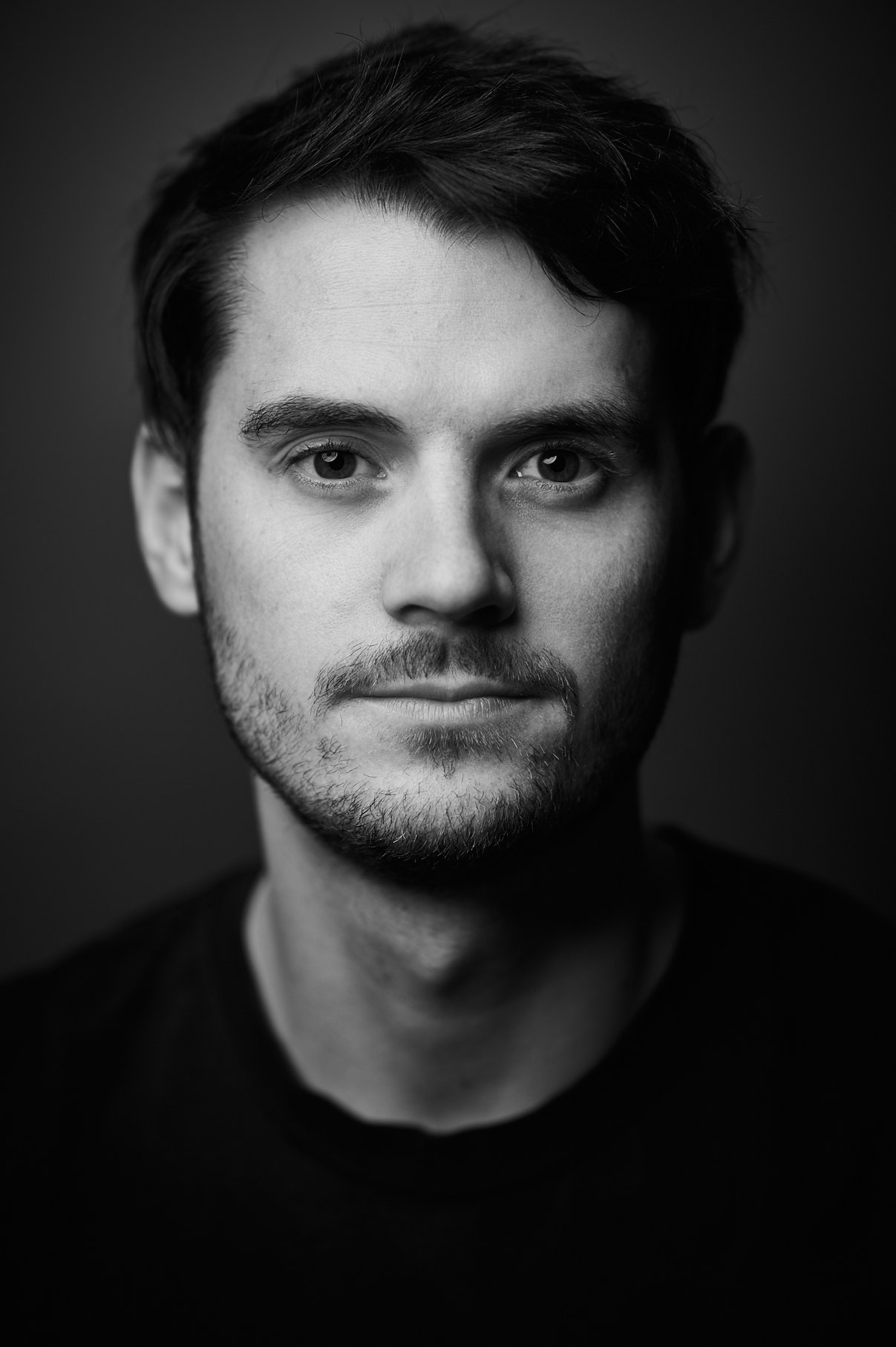}}]{Maximilian Stahlke} received his Master’s degree in Electronic and Mechatronic Systems at the Institute of Technology Georg Simon Ohm, Germany, in 2020. Since 2020 he works at the precise positioning and analytics department at Fraunhofer IIS in the Hybrid Positioning \& Information Fusion group. His research interests are hybrid positioning for radio-based localization systems with the focus on model- and data-driven information fusion.\end{IEEEbiography}

\vspace{-1cm}\begin{IEEEbiography}
[{\includegraphics[width=1in,height=1.25in,clip,keepaspectratio]{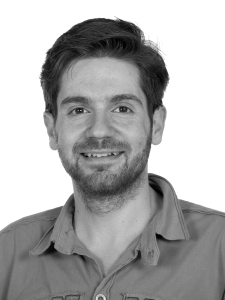}}]{George Yammine} received the B.E.\ degree in 2010 in computer and communications engineering from Notre Dame University, Louaize, Lebanon, the M.Sc.\ in communications technology and the Dr.-Ing.\ degrees from Ulm University, Ulm, Germany, in 2014 and 2020, respectively. His dissertation was on noncoherent detection in massive MIMO systems. From 2005 until 2020, he was a Research Assistant at the Institute of Communications Engineering, Ulm University. 

In 2021, he joined the Positioning in Wireless Networks group at Fraunhofer Institute for Integrated Circuits IIS, N\"urnberg, Germany. Currently, he is a Senior Scientist working on different aspects of 6G mobile communications and positioning. His research interests include noncoherent detection techniques, signal processing, advanced positioning techniques, multi-user detection, machine learning applied in communications and signal processing, reconfigurable intelligent surfaces and massive MIMO systems. 

Dr.\ Yammine was the recipient of the ARGUS Science Award 2014 from Airbus Defence and Space.
\end{IEEEbiography}

\vfill

\vspace{-1cm}\begin{IEEEbiography}[{\includegraphics[width=1in,height=1.25in,clip,keepaspectratio]{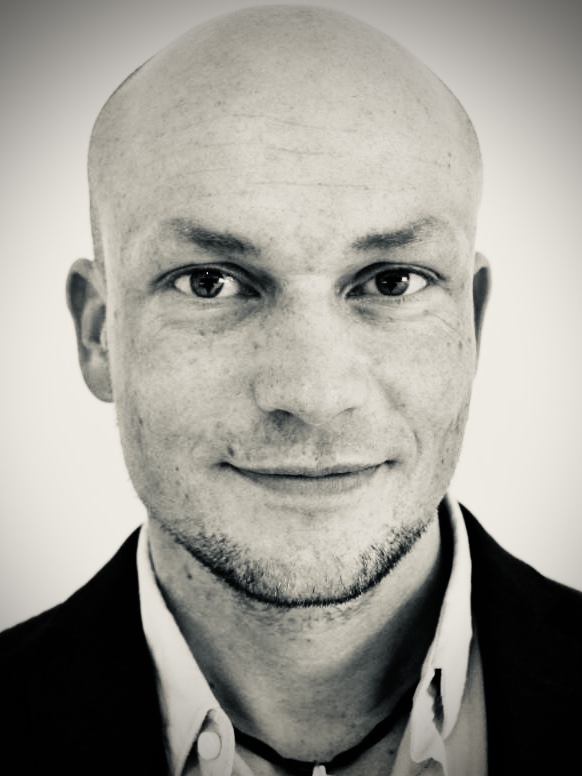}}]{Tobias Feigl} received his Ph.D.\ degree in Computer Science from the Friedrich-Alexander-University Erlangen-Nuremberg (FAU) in 2021 and his Masters degree from the University of Applied Sciences Erlangen-Nuremberg, Germany, in 2017. In 2009 he joined the Fraunhofer Institute for Integrated Circuits (IIS) Nuremberg. 
In parallel, since 2017 he is a lecturer at the Computer Science department (Programming Systems lab) at FAU, where he gives courses on machine and deep learning and supervises related qualification work. 
His research combines human-computer interaction and machine learning to improve localization. \end{IEEEbiography}

\vspace{-1cm}\begin{IEEEbiography}[{\includegraphics[width=1in,height=1.25in,clip,keepaspectratio]{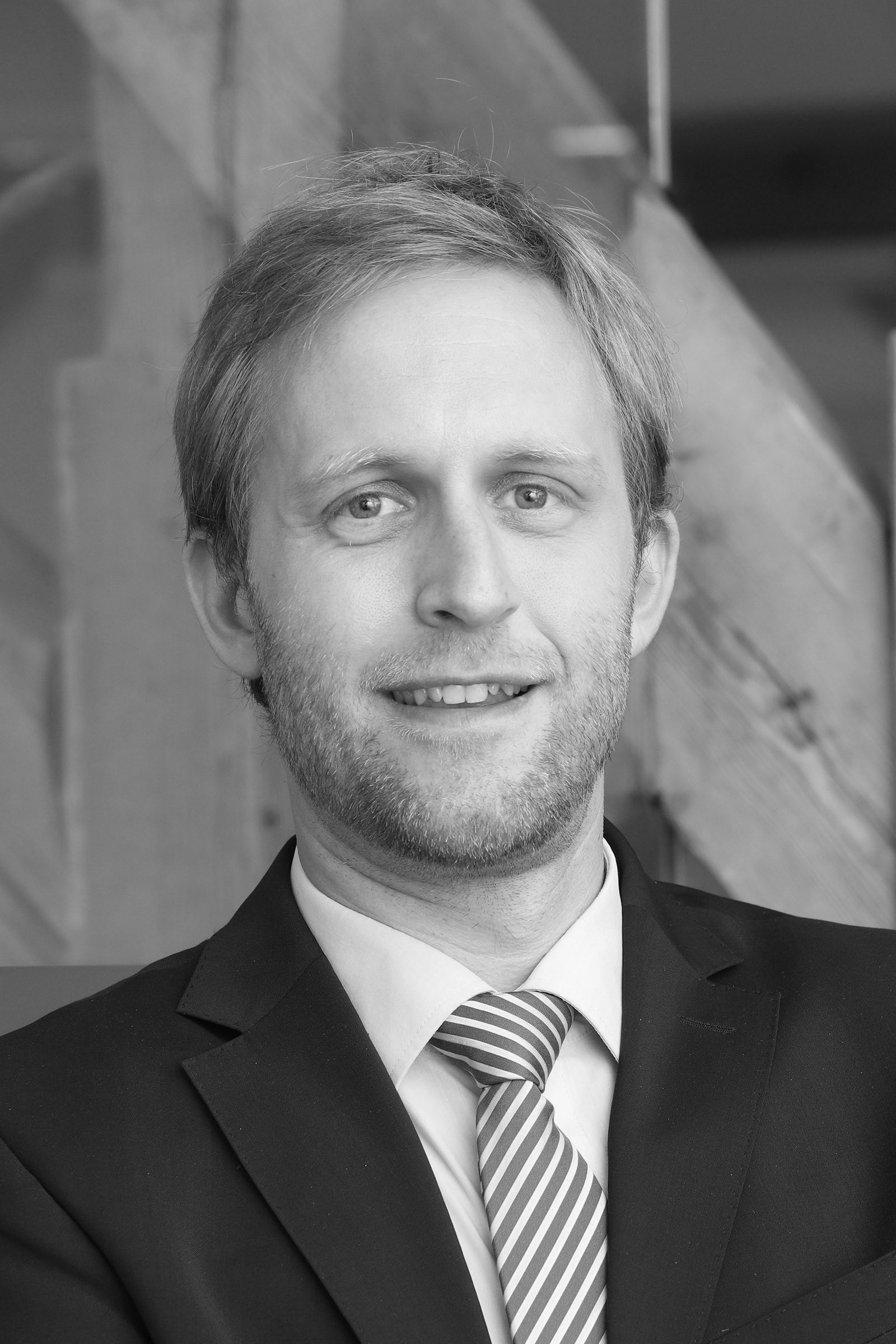}}]{Bjoern M.\ Eskofier} heads the Machine Learning and Data Analytics (MaD) Lab at the Friedrich-Alexander-University Erlangen-N\"urnberg (FAU). He is also the current speaker of FAU’s Department Artificial Intelligence in Biomedical Engineering (AIBE) and of the German Ministry of Economic Affairs and Climate Action GAIA-X usecase project “TEAM-X” and the co-speaker of the German Research Foundation collaborative research center “EmpkinS” (SFB 1483).

Dr.\ Eskofier studied Electrical Engineering at the FAU and graduated in 2006. He then studied under the supervision of Prof.\ Dr.\ Benno Nigg at the University of Calgary (Canada). He authored more than 300 peer reviewed articles, submitted 7 patent applications, started three spinoff startup companies, and is in a supporting role for further startups. He won several medical-technical research awards, including the “Curious Minds” award in “Life Sciences” by Manager Magazin and Merck. In 2016, he was a visiting professor in Dr.\ Paolo Bonato’s Motion Analysis Lab at Harvard Medical School (February-March), and in 2018, he was a visiting professor in Dr.\ Alex “Sandy” Pentland’s Human Dynamics group at MIT Media Lab (March-August).

Bjoern Eskofier has defined his research and entrepreneurial agenda to revolve around contributions to a ``Digital Health Ecosystem'', where patients are connected to other stakeholders within the Healthcare system using digital support tools. His digital health research philosophy is that only multidisciplinary teams of engineers, medical experts, industry representatives and entrepreneurs will have the tools to actually implement changes in Healthcare.
\end{IEEEbiography}

\vspace{-1cm}\begin{IEEEbiography}[{\includegraphics[width=1in,height=1.25in,clip,keepaspectratio]{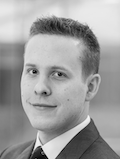}}]{Christopher Mutschler} leads the precise positioning and analytics department at Fraunhofer IIS. Prior to that, Christopher headed the Machine Learning \& Information Fusion group. He gives lectures on machine learning at the FAU Erlangen-N{\"u}rnberg (FAU), from which he also received both his Diploma and Ph.D.\ in 2010 and 2014, respectively. Christopher’s research combines machine learning with radio-based localization.\end{IEEEbiography}

\vfill

\end{document}